\documentclass{article}

\usepackage{arxiv}

\usepackage[utf8]{inputenc} 
\usepackage[T1]{fontenc}    
\usepackage{hyperref}       
\usepackage{url}            
\usepackage{booktabs}       
\usepackage{amsfonts}       
\usepackage{nicefrac}       
\usepackage{microtype}      
\usepackage{lipsum}

\usepackage{epsfig}
\usepackage[displaymath]{lineno}
\usepackage{hyperref}
\usepackage{graphicx,pstricks}
\usepackage{graphics}
\usepackage{amssymb}
\usepackage{amsmath, amsthm}
\usepackage{algorithm}
\usepackage{booktabs,tabu,multirow}
\usepackage{subfig}
\usepackage{caption}
\usepackage{colortbl}
\usepackage{setspace}
\usepackage{bbm}

\title{Effective Data Sampling Strategies and Boundary Condition Constraints of Physics-Informed Neural Networks for Identifying Material Properties in Solid Mechanics}

\author{
\textbf{Wensi Wu}\\
Department of Anesthesiology and\\
Critical Care Medicine,\\
Division of Pediatric Cardiology,\\
Children's Hospital of Philadelphia,\\
Philadelphia, PA 19104 \\
\texttt{wuw4@chop.edu} \\
\And
\textbf{Mitchell Daneker}\\
Department of Chemical and \\
Biomolecular Engineering,\\ 
University of Pennsylvania,\\
Philadelphia, PA 19104\\
\texttt{mdaneker@seas.upenn.edu} \\ 
\And
\textbf{Matthew A. Jolley}\\
Department of Anesthesiology and\\ 
Critical Care Medicine,\\
Division of Pediatric Cardiology,\\
Children's Hospital of Philadelphia,\\
Philadelphia, PA 19104\\
\texttt{jolleym@chop.edu} \\  
\And
\textbf{Kevin T. Turner}\\
Department of Mechanical Engineering\\ 
and Applied Mechanics,\\
University of Pennsylvania,\\
Philadelphia, PA 19104\\
\texttt{kturner@seas.upenn.edu} \\ 
\And
\textbf{Lu Lu}\\
Department of Chemical and \\
Biomolecular Engineering,\\ 
University of Pennsylvania,\\
Philadelphia, PA 19104\\
\texttt{lulu1@seas.upenn.edu} \\  
}

\begin{document}
\maketitle

\begin{abstract}
Material identification is critical for understanding the relationship between mechanical properties and the associated mechanical functions. However, material identification is a challenging task, especially when the characteristic of the material is highly nonlinear in nature, as is common in biological tissue. In this work, we identify unknown material properties in continuum solid mechanics via physics-informed neural networks (PINNs). To improve the accuracy and efficiency of PINNs, we developed efficient strategies to nonuniformly sample observational data. We also investigated different approaches to enforce Dirichlet boundary conditions as soft or hard constraints. Finally, we apply the proposed methods to a diverse set of time-dependent and time-independent solid mechanic examples that span linear elastic and hyperelastic material space. The estimated material parameters achieve relative errors of less than 1\%. As such, this work is relevant to diverse applications, including optimizing structural integrity and developing novel materials.
\end{abstract}



\section{Introduction}

The study of nonlinear dynamical systems is of interest in many science and engineering disciplines due to the nonlinear nature of most physical phenomena. The nonlinear relation between system inputs and outputs and interactions between system variables have made nonlinear systems a daunting task to solve with traditional analytical approaches. In engineering, the solutions to nonlinear dynamical systems are approximated by classic numerical methods (\textit{i.e.,} finite difference method, finite element method, or finite volume method). These numerical methods discretize a large system into finite spatial subcomponents, linearize the governing equations in time, and solve the linearized equations iteratively until each subcomponent satisfies the governing conservation laws. However, this spatio-temporal discretization procedure often introduces spurious oscillation and requires numerical damping in order to obtain stable solutions, which may lead to less accurate approximations~\cite{newmark1959, hilber1977, chung1993, bruls2008}. On the contrary, the universal approximation theorem of neural networks \cite{hornik1989} states that a sufficiently large feed-forward artificial neural network with proper nonlinear activation functions can approximate any continuous function. As such, machine learning-based approaches have arisen as a new paradigm for addressing physical problems that are known to be challenging for classic numerical methods~\cite{karniadakis2021physics}. These recent advances in deep learning techniques have demonstrated considerable potential for unveiling the hidden physics of many complex nonlinear dynamical systems where classic numerical methods fail \cite{raissi2018, mao2020, cai2021,karniadakis2021physics,lu2021learning}.

The tremendous growth of deep learning has also attracted material scientists' attention to accelerating the understanding of complex material properties. A comprehensive understanding of mechanical properties is essential for studying the behaviors of materials under load. In the traditional engineering approach, the material investigation (material parameter estimation) procedure is generally carried out in the following three steps: 1) conduct a series of experimental studies to quantify the mechanical behavior of the material; 2) identify a representative mathematical model by performing a series of inverse finite element analyses; and 3) use optimization techniques to identify the unknown parameters in the mathematical model that produce the best agreements with experimental results. However, the inversion process can be computationally expensive, or even impossible, for complex, nonlinear models owing to a large number of forward simulations required \cite{guo2021}. As a result, data-driven deep learning has been leveraged as a surrogate modeling technique to study the nonlinear deformation relationship between material behavior and load conditions \cite{chen2019, wu2020, stern2020,lu2020, linka2021}, as well as to designing new materials with unique mechanical characteristics~\cite{xue2016, conduit2017, ling2017, spear2018, gu2018, kim2021}.

Despite the efficacy of deep learning in interpreting complex systems, this promising method is not without shortcomings. First, the accuracy of deep learning predictions is highly reliant on the volume of data~\cite{jin2020quantifying}. Second, conventional neural networks trained purely on data are unrestricted to the system's underlying governing equations and boundary conditions; this limits the capability to extrapolate accurate physical relations from network outputs beyond their training data \cite{linka2021}. As a solution to this limitation, physics-informed neural networks (PINNs)~\cite{raissi2019,lu2021} have emerged to improve the training process by integrating mathematical models. PINNs use neural networks to approximate the solution and encode the governing equations (\textit{e.g.,} the ordinary differential equations or partial differential equations) in the loss function. For an inverse problem, this loss function encompasses the residual of the initial condition, boundary conditions, the PDE at specific collocation points in the physical domain, and observation data. Incorporating physical laws ensures the networks satisfy both phenomenological observations from data and the underlying physical laws and constraints within the system, and therefore significantly improve the effectiveness of applying the trained models to unexplored data sets and for sensitivity analysis \cite{haghighat2021}. For example, Lu et al. demonstrated that integrating physical laws and experimental data results in significantly improved solution accuracy for extracting material properties from instrumented indentation tests \cite{lu2020}.

PINNs have succeeded in estimating the solutions to a wide range of forward and inverse problems, including classic differential problems \cite{raissi2019, chen2020, mao2020, jin2021, almajid2022, cuomo2022,yazdani2020systems,daneker2022systems}, fractional equations \cite{pang2019}, integral-differential equations \cite{yuan2022, lu2021}, and stochastic partial differential equations (PDEs) \cite{zhang2019}. In recent years, researchers have used PINNs to address nonlinear solid mechanic problems by modifying the network architecture, the loss function expression~\cite{yu2022gradient}, and collocation point sampling methods. Samaniego et al.~\cite{samaniego2020}, Nguyen-Thanh et al.~\cite{nguyen2020, nguyen2021}, and Abueidda et al.~\cite{abueidda2021} developed a deep energy method (\textit {i.e.,} the loss term for the PDEs was expressed in terms of the potential energy rather than the conservation of momentum) and demonstrated its applicability to steady state elasticity, hyperelasticity, viscoelasticity, and piezoelectricity problems. Fugh et al. discovered that incorporating stresses as additional outputs in the network enhances the network's capability of resolving localized features in the solution for linear elastic steady state problems~\cite{fuhg2022}. Haghighat et al. presented a parallel neural network architecture to identify material parameters for linear elastic and nonlinear-elastoplastic test problems using a pre-trained network~\cite{haghighat2021}. Henkes et al. presented an adaptive collocation sampling points framework to capture the underlying physics of microstructural elastostatics \cite{henkes2022}. Wu et al. proposed residual-based adaptive sampling methods and demonstrated significantly improved prediction accuracy for both forward and inverse problems \cite{wu2023}. Rao et al. presented a PINN framework for solving forward elastic and elastodynamic problems with strongly enforced initial conditions and boundary conditions. Zhang et al. presented a PINN framework to identify unknown geometric and material parameters of steady state linear elastic, hyperelastic, and plastic materials with a pre-trained network \cite{zhang2022}. 

In the present work, we aim to derive unknown material parameters in continuum solid mechanics. We focus on the novel application of PINNs to identify the unknown material parameters on higher-order initial value and boundary value problems. We identify the performance of enforcing Dirichlet boundary conditions as soft or hard constraints. We also show the efficacy of several observation point sampling strategies on the 2D examples for estimating linear elastic and hyperelastic materials. For the static problems, we considered 1) concentrated sampling in the location with high stress differential, 2) uniform sampling across the spatial domain, and 3) sampling along the boundary only. For the dynamic example, we experimented with reducing the number of time frames in the reference data. 

Applying PINNs to inverse problems allows the discovery of material constitutive properties in a wide range of engineering domains when they are difficult, or impossible, to obtain otherwise. This work demonstrates a proof of concept of applying PINNs to uncover unknown material constants in test examples under various material types and loading conditions. We demonstrate the applicability of our PINN inversion framework to both steady state and dynamic solid mechanics examples by applying our methods to the wave equation, Euler-Bernoulli beam equation, and solid mechanics momentum conservation equations. The estimated parameters are within 1\% of relative errors compared against the true values in 4 out of 5 test examples and within 2.5\% in all examples. The excellent prediction accuracy in our work indicates a promising framework for improving engineering system performance and material designs.

This paper is organized as follows. In Section \ref{sec:methods}, we first introduce the geometry and governing equations for the classic solid mechanic examples in the present work; we then present an overview of neural network architectures and loss function setup for both soft and hard constraints; after that, we describe the choice of auxiliary functions for hard constraints; at the end, we delineate additional technical details germane to achieving high prediction accuracy in solid mechanics problems. In Section \ref{results}, we first demonstrate the effectiveness of various data point sampling strategies. We then compare the performance of soft and hard constraints. Subsequently, we present parameter estimation results for both 1D and 2D examples. Finally, in Sections \ref{discussion} and \ref{conclusion}, we summarize the novelty and benefits of the proposed methods and conclude the paper.

\section{Methods}
\label{sec:methods}

In this work, we apply PINNs to estimate material parameters of solid mechanics problems. In addition, we provide details on the neural network architectures, loss functions, boundary condition constraints, and technical considerations specific to solving inverse problems in solid mechanics.

\subsection{Test examples}
\label{sec:examples}
We consider five classic examples to cover a spectrum of steady state and dynamic analyses in 1D and 2D space, representing partial differential equations of up to fourth order in space and second order in time. The 1D examples are governed by the wave and the Euler-Bernoulli beam equation, respectively. The 2D examples are governed by the conservation of momentum, the material constitutive model, and the kinematic relation. The governing equations for each test example are detailed in Section \ref{govern_eqns}. 

\begin{figure}[htbp]
\centering
\includegraphics[width=1\textwidth]{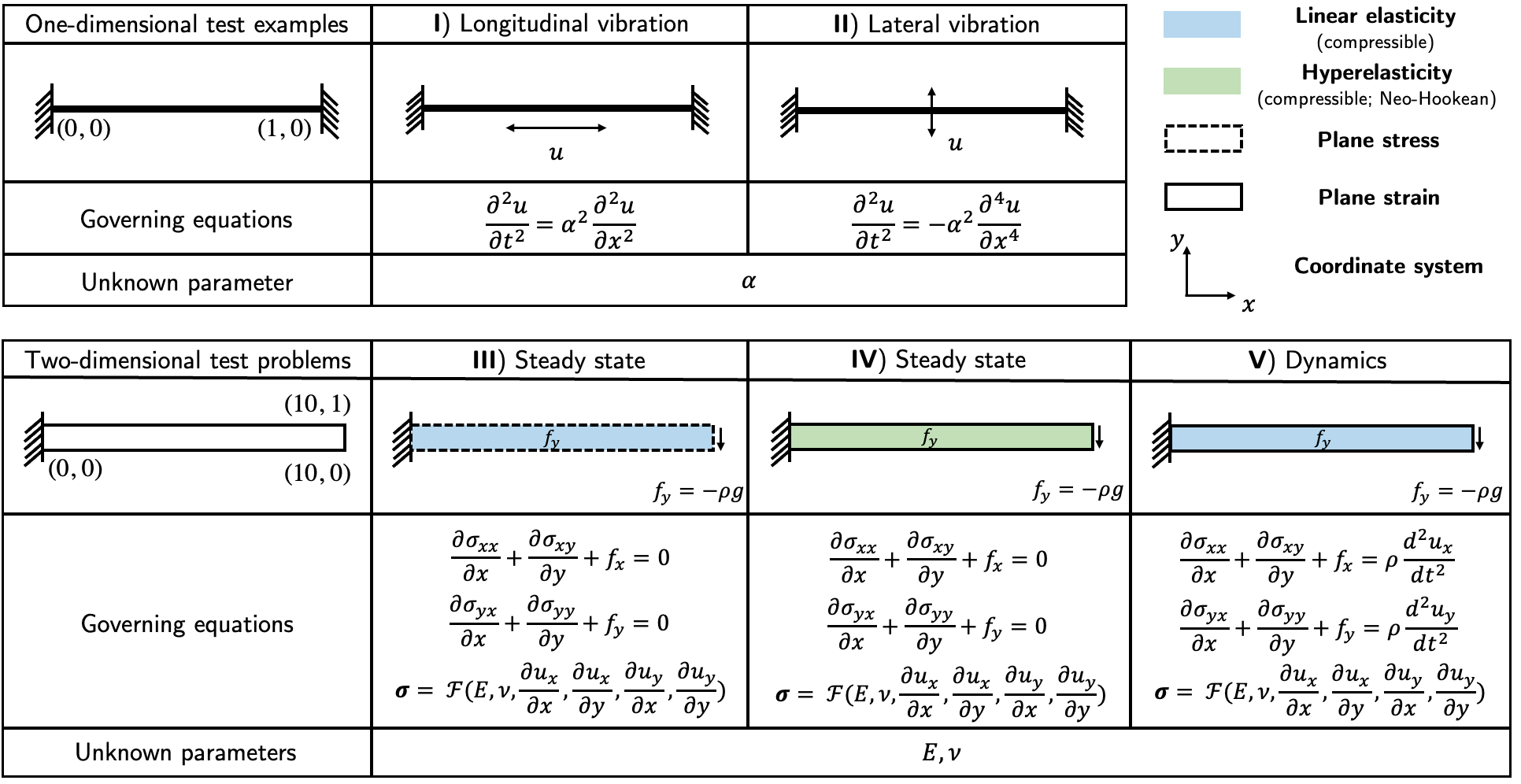}
\caption{\textbf{Five classic solid mechanics examples.} The geometry of the examples is presented in the undeformed configuration. In addition, we provide the material model, stress configuration, and loading condition for each problem. The aim is to identify unknown parameters $\alpha$ in the one-dimensional test examples and unknown parameters $E$ and $\nu$ in the two-dimensional test examples. A detailed description of the governing equations is presented in Section \ref{govern_eqns}.}\label{overview}
\end{figure}

In these inverse problems, we identify the material parameters by utilizing the underlying governing equations and deformation/stress data obtained from either analytical solution or finite element analysis. We provide finite element analysis verficatioin results in Section~\ref{verfication}. To obtain reference data for training, in the 1D examples, we consider a beam that is fixed on both ends. We compute the deformation training data from the analytical solution by applying longitudinal/lateral initial deformation to the beam. In the 2D examples, we consider a cantilever beam that is fixed on the left end. We apply a body force to the structure and use finite element analysis to compute the displacements and stresses on the beam. The details of the model geometry, material model (incompressible linear elastic or compressible hyperelastic material), stress configuration (plane stress or plane strain), and loading condition (applied deformation or body force) are specified in Fig.~\ref{overview}.

In the following, we present the governing equations for a 2D elastodynamics analysis. The momentum equation is expressed as: 
\begin{equation*}
\sigma_{ij,j}+f_i=\rho \partial_{tt}u_{i}, 
\end{equation*}
where $\rho$ is the material density, $f$ is the externally applied force, and the subscript comma denotes partial derivatives. The isotropic linear elastic material constitutive model with plane-strain is:
\begin{equation*}
\sigma_{ij} = \frac{E \nu}{(1+\nu)(1-2\nu)} \delta_{ij} \epsilon_{kk}+ \frac{E}{(1+\nu)} \epsilon_{ij}.
\end{equation*}
Herein, $E$ and $\nu$ are Young's modulus and Poisson's ratio to be estimated using PINNs, $\delta_{ij}$ is the Kronecker delta, and $\epsilon_{ij}$ is the infinitesimal strain tensor. Lastly, the kinematic relation is written as:
\begin{equation*}
\epsilon_{ij} = \frac{1}{2}(u_{i,j}+u_{j,i}).
\end{equation*}

\subsection{Physics-informed neural networks for solid mechanics} \label{pinns_formulation}
Here, we provide an overview of the PINN formulation for solving solid mechanics inverse problems. The inverse PINN framework is set up using the DeepXDE library \cite{lu2021}, and the code is publicly available from the GitHub repository \url{https://github.com/lu-group/pinn-material-identification}.

\subsubsection{Neural network architectures}

Let $\mathcal{N}^{L}(\mathbf{x}) \colon \mathbb{R}^{\dim(\mathbf{x})} \to \mathbb{R}^{\dim(\mathbf{y})}$, be a $L$-layer neural network that maps input features $\mathbf{x}$ to output variables $\mathbf{y}$ with $\mathcal{N}^l$ neurons in the $l$-layer. The connectivity between layer $l$ and $l-1$ is governed by $\mathcal{N}^l = \phi(\mathbf{W}^l\mathcal{N}^{l-1}(\mathbf{x})+\mathbf{b}^l)$, where $\phi$ is a nonlinear activation function, $\mathbf{W}^{l}$ is a weight matrix, and $\mathbf{b}^l$ is a bias vector. We use hyperbolic tangent, $\tanh$, as the activation function in the present work. Given that the activation function is applied element-wise to each neuron, the recursive FNN is defined as:
\begin{align*}
\textrm{input layer}: \quad & \mathcal{N}^0(\mathbf{x}) = \mathbf{x} \in \mathbb{R}^{\dim{\mathbf{x}}}, \\
\textrm{hidden layer $l$}: \quad & \mathcal{N}^l(\mathbf{x}) = \tanh \left(\mathbf{W}^l\mathcal{N}^{l-1}(\mathbf{x})+\mathbf{b}^l \right) \in \mathbb{R}^{\mathcal{N}^l}, \quad \textrm {for} \quad 1\leq l \leq L-1, \\
\textrm{output layer}: \quad & \mathcal{N}^L(\mathbf{x}) = \mathbf{W}^L\mathcal{N}^{L-1}(\mathbf{x})+\mathbf{b}^L \in \mathbb{R}^{\dim{(\mathbf{y})}}.
\end{align*}

\begin{figure}[htbp]
\centering
\includegraphics[width=1\textwidth]{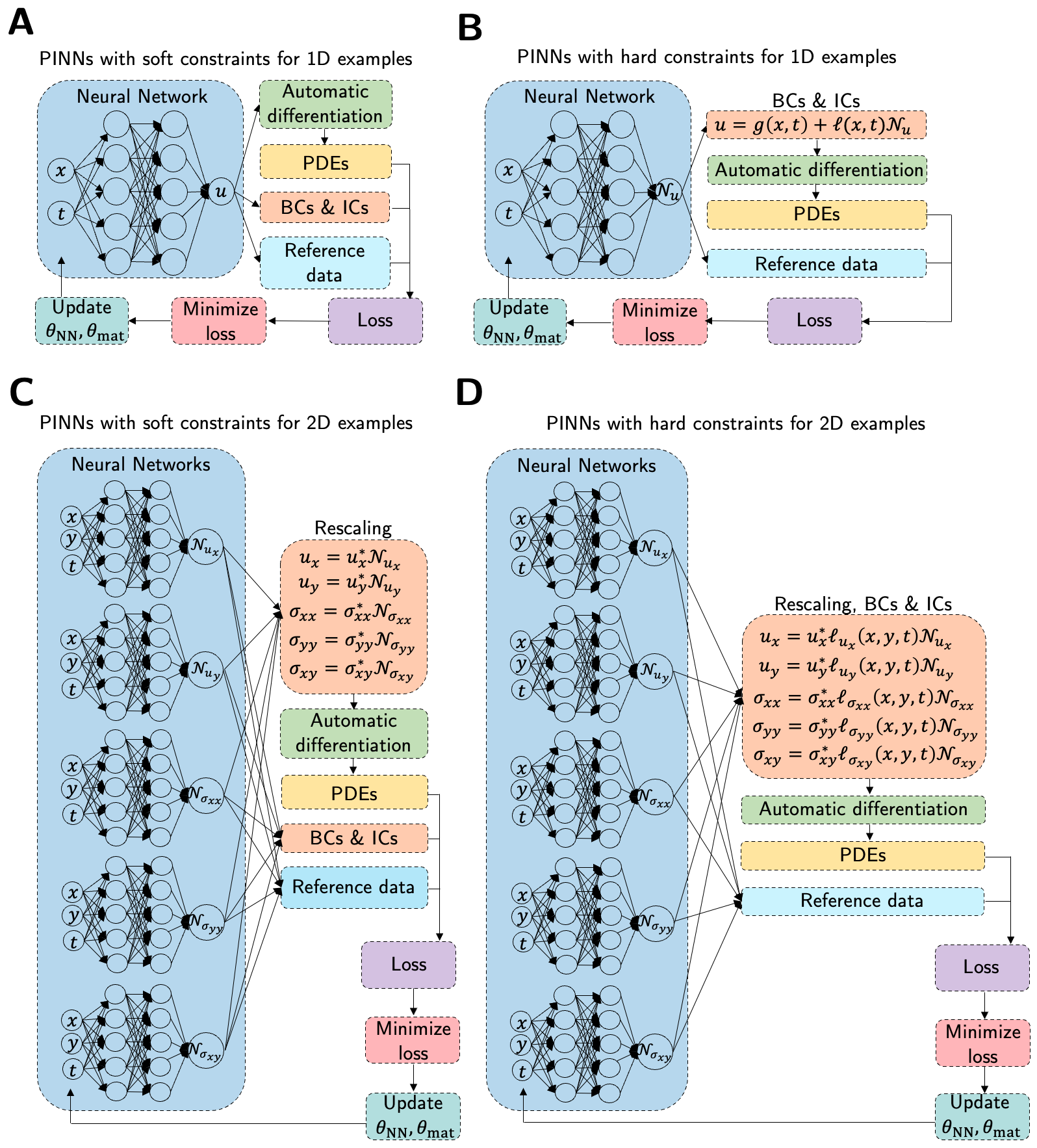}
\caption{\textbf{PINN architectures for nonlinear solid mechanics systems.} We use one FNN for the 1D examples (\textbf{A} and \textbf{B}) and five independent FNNs for the 2D examples (\textbf{C} and \textbf{D}). We consider soft and hard constraints for both 1D and 2D examples.}\label{architecture}
\end{figure}

The PINN architectures are presented in Fig. \ref{architecture}. In the 1D examples (Fig. \ref{architecture}A and B), \{$x$, $t$\} are network input features corresponding to x- Cartesian coordinate and time, respectively, and \{$u$\} is the predicted displacement. In the 2D examples (Figs. \ref{architecture}C and D), \{$x$, $y$, $t$\} are network input features corresponding to x- and y- Cartesian coordinates, and time, respectively. Note that the steady state cases only have two input features: x- and y- Cartesian coordinates. \{$\mathcal{N}_{u_x}$, $\mathcal{N}_{u_y}$, $\mathcal{N}_{\sigma_{xx}}$, $\mathcal{N}_{\sigma_{yy}}$, $\mathcal{N}_{\sigma_{xy}}$\} are network output. \{$u_x$, $u_y$, $\sigma_{xx}$, $\sigma_{yy}$, $\sigma_{xy}$\} are the predicted displacements and stresses. In all four architectures, \{$\theta_\text{NN}$\} represents network parameters ($\mathbf{W}^{l}$ and $\mathbf{b}^l$); \{$\theta_\text{mat}$\} represents unknown material variables ($\alpha$ for 1D, and $E$ and $\nu$ for 2D).

For complex problems, the network needs to be sufficiently large in order to capture the cross-dependence between variables in the governing equations. Aside from creating a single large network, an alternative approach is establishing a separate, independent network for each output variable. In the present work, we model the test examples using multiple independent networks, given their effectiveness over the single network approach \cite{haghighat2021}. In cases with only one output variable, the network architecture reduces to a standard single fully-connected neural network (FNN). In other words, one FNN is used in the time-dependent 1D vibration test cases, and five independent parallel FNNs are used in the 2D cantilever beam test cases.

We briefly summarize the training procedure of both architectures (Fig.~\ref{architecture}). We first initialize network parameters ($\theta_{\text{NN}}$) to obtain an initial estimation of displacements and/or stresses. Given the initial estimation, we subsequently formulate a loss function to minimize the PDE residuals, as well as the errors between the ground truth and the approximation in the boundary, the initial domain, and reference data points (also known as observational data points). The network variables ($\theta_\text{NN}$, $\theta_\text{mat}$) are then updated iteratively until the total loss converges, and the estimated material constants reach a plateau.

\subsubsection{Loss functions}

In the training process, we optimize the network parameters, $\theta_{\text{NN}}$ (\textit{i.e.,} $\mathbf{W}^l$ and $\mathbf{b}^l$), and the unknown material parameters, $\theta_{\text{mat}}$ (\textit{i.e.,} $E$ and $\nu$), expressed as:
\begin{equation*}
\theta_{\text{NN}}^*, \theta_{\text{mat}}^* = \underset{\theta_{\text{NN}}, \theta_{\text{mat}}}{\arg\min} \mathcal{L}(\theta_{\text{NN}}, \theta_{\text{mat}}),
\end{equation*}
where $\mathcal{L}(\theta_{\text{NN}}, \theta_{\text{mat}})$ is the loss function that measures the total error between network outputs concerning the model's initial conditions, boundary conditions, PDE evaluations (conservation of momentum and constitutive material laws) on the collocation points, and the reference data. The total loss function is defined as: 
\begin{equation*}
\mathcal{L}(\theta_{\text{NN}}, \theta_{\text{mat}}) =
w_{\text{BCs}}\mathcal{L}_{\text{BCs}} +
w_{\text{ICs}}\mathcal{L}_{\text{ICs}} +  w_{\text{Data}}\mathcal{L}_{\text{Data}} + w_{\text{PDEs}}\mathcal{L}_{\text{PDEs}}, 
\end{equation*}
where $w_\bullet$ is the weight associated with its corresponding loss term $\mathcal{L}_\bullet$. The loss terms $\mathcal{L}_{\text{BCs}}$, $\mathcal{L}_{\text{ICs}}$, $\mathcal{L}_{\text{Data}}$ compute the mean squared error of predicted results on the collocation points for the boundary, initial configuration, and observation data, respectively. Further, $\mathcal{L}_{\text{PDEs}}$ computes the mean squared error of the PDE residuals over the spatial-temporal domain.

In the time-dependent longitudinal and lateral vibration examples, the PDE loss constitutes the residual in the wave equation and Euler-Bernoulli beam equation, respectively. In the 2D examples, the PDE loss comprises the residuals in the constitutive relations and momentum conservation laws. Detailed formulations of the material models and the governing PDEs are presented in Section~\ref{govern_eqns}. The loss function formulation varies slightly depending on the nature of the example problems (\textit{i.e.,} steady state or dynamic). For a 2D linear elastic dynamic example, $\mathcal{L}_{\text{PDEs}}$ is formulated as: 
\begin{align*}
\mathcal{L}_{\text{PDEs}} = & \|\frac{\partial \sigma_{xx}}{\partial x} +\frac{\partial \sigma_{xy}}{\partial y}+f_x-\rho\frac{\partial^2u_x}{\partial t^2}\|^2_2 \\
+ & \|\frac{\partial \sigma_{xy}}{\partial y}+\frac{\partial \sigma_{yy}}{\partial y}+f_y-\rho\frac{\partial^2u_y}{\partial t^2}\|^2_2 \\
+ & \|\sigma_{xx} - \frac{E_{\text{pred}}}{(1+\nu_{\text{pred}})(1-2\nu_{\text{pred}})}\left[(1-\nu_{\text{pred}})\epsilon_{xx}+\nu_{\text{pred}}\epsilon_{yy}\right]\|^2_2 \\
+ & \|\sigma_{yy} - \frac{E_{\text{pred}}}{(1+\nu_{\text{pred}})(1-2\nu_{\text{pred}})}\left[(1-\nu_{\text{pred}})\epsilon_{yy}+\nu_{\text{pred}}\epsilon_{xx}\right]\|^2_2 \\
+ & \|\sigma_{xy} - \frac{E_{\text{pred}}}{(1+\nu_{\text{pred}})}\epsilon_{xy}\|^2_2.
\end{align*}

In order to compute the PDE loss $\mathcal{L}_{\text{PDEs}}$,  we require higher order derivatives of the network output (\textit{i.e.,} the displacements) both in time and space. The partial derivatives in the governing equations are approximated using a technique called automatic differentiation. This method evaluates the derivatives by applying the chain rule during back-propagation \cite{rumelhart1986, lu2021}. This approach has been shown to be more efficient than conventional numerical methods for estimating derivatives (\textit{i.e.,} finite difference, symbolic differentiation, etc.)~\cite{baydin2017, margossian2019}.

In some cases, we consider hard constraint initial and boundary conditions. Meaning, we impose the initial and boundary conditions on the network outputs before evaluating the governing equations \cite{lu2021_2}. More information on hard constraints will be discussed in Section~\ref{sec:hard_bc}. As such, the loss function is simplified into two loss terms:
\begin{equation*}
\mathcal{L}(\theta_{\text{NN}}, \theta_{\text{mat}}) =
 w_{\text{Data}}\mathcal{L}_{\text{Data}} +
 w_{\text{PDEs}}\mathcal{L}_{\text{PDEs}}. 
\end{equation*}
The values of the weights for each example are in Section~\ref{govern_eqns}.

\subsection{Hard constraint initial and boundary conditions}
\label{sec:hard_bc}

In PINNs, initial and boundary conditions are usually weakly imposed as soft constraints due to simplicity. However, soft constraints do not guarantee the approximate solution to satisfy the initial and boundary conditions, which can affect the accuracy of the parameter prediction. Hence, we also consider imposing initial and Dirichlet boundary conditions via hard constraints \cite{lu2021_2} to ensure the PDEs satisfy the initial and Dirichlet boundary conditions exactly.

Let us consider $\mathcal{N}(\mathbf{x})$ to be the network output, and we aim to satisfy the initial and Dirichlet boundary conditions such that:
\begin{equation*}
\mathcal{N}^\prime(\mathbf{x}) = g_0(\mathbf{x}), \quad \mathbf{x} \in \Gamma_D \cup \Omega_{0}, 
\end{equation*}
where $\Gamma_D \subset \partial \Omega$ is a subset of the boundary and $\Omega_0$ is the initial domain. The hard constraint is achieved by using two auxiliary functions $g(\mathbf{x})$ and $\ell(\mathbf{x})$ (Figs.~\ref{architecture}B and D) such that:
\begin{equation*}
\mathcal{N}^\prime(\mathbf{x}) = g(\mathbf{x}) + \ell(\mathbf{x})\mathcal{N}(\mathbf{x}).
\end{equation*}
Here, $g(\mathbf{x})$ is a function that satisfies the required initial and boundary conditions (which could be either zero or non-zero). Further, $\ell(\mathbf{x})$ is a function that satisfies the following conditions: 
\begin{gather*}
 \begin{cases}
   \ell(\mathbf{x}) = 0 , & \mathbf{x} \in \Gamma_D \cup \Omega_{0},\\
    \ell(\mathbf{x}) > 0,  & \mathbf{x} \in \Omega \setminus (\Gamma_D \cup \Omega_{0}).
\end{cases}
\end{gather*}
Interested readers can refer to this \cite{lu2021_2} for more information on defining suitable auxiliary functions.

In this study, we compare the performance of soft constraints and hard constraints. The choice of $\ell(\mathbf{x})$ is not unique, and based on our experience, it influences the prediction accuracy. For instance, in the 2D examples, we consider a continuous and a discontinuous function for $\ell({\mathbf{x}})$ and compare their performance. For the continuous function, we simply choose $g(\mathbf{x}) = 0$ and $\ell(\mathbf{x}) = x$. For the discontinuous function, we choose $g(\mathbf{x}) = 0$ and $\ell(\mathbf{x}) = \mathbbm{1}_{\{x>0\}}$. Both options yield satisfactory results, but to our surprise, the latter offers better material parameter prediction; this indicates that further research is required to better understand the effect of the auxiliary function characteristics on prediction accuracy.

\subsection{Additional technical details}

In many solid mechanics problems, the displacements and stresses are often several orders of magnitude different from $\mathcal{O}(1)$. This wide spread of magnitude orders can present a challenge for the network training process. In the 2D examples with $g(\mathbf{x}) = 0$, to achieve an accurate solution, we rescale the predicated displacement and stresses by multiplying their corresponding maximum absolute values obtained from the observational data. After that, we apply hard constraints to the variables, such that the magnitude of the network output is of $\mathcal{O}(1)$: 
\begin{gather*}
u_i = u^*_{i}\ell(\mathbf{x})\mathcal{N}_{u_i}, \\
\sigma_{ij} = \sigma^*_{ij}\ell(\mathbf{x})\mathcal{N}_{\sigma_{ij}}.
\end{gather*}
Here, \{$u^*_x$, $u^*_y$, $\sigma^*_{xx}$, $\sigma^*_{yy}$, $\sigma^*_{xy}$\} (Figs.~\ref{architecture}C and D) are the maximum absolute displacements and stresses from observational data used for variable rescaling.

When approximating unknown variables, the estimated values can converge to a local minimum or trivial solution that satisfies the governing equations but are unrealistic in the physical system. As such, we use a $\tanh$ function to constrain the approximation to a realistic range of values. This modification helps guide the network toward meaningful estimations. In particular, for the 2D dynamic example, we set
\begin{gather*}
    E_{\text{pred}} = 5\times 10^6\times[\tanh(\hat{E}_{\text{pred}})+1] \quad \text{and} \quad  \nu_{\text{pred}} = \frac{1}{4} [\tanh(\hat{\nu}_{\text{pred}})+1]
\end{gather*}
where $\hat{\bullet}_{\text{pred}}$ is an auxiliary variable and $\bullet_{\text{pred}}$ is a scaled variable for use in the PDE calculation. This conversion ensures that all network predicted auxiliary variables are of a similar scale, which helps improve solution convergence.

\section{Results}
\label{results}

Before we apply PINNs to the sample problems, we first examine suitable observation point sampling and boundary constraint methods, detailed in Sections \ref{re:observation_points} and \ref{re:boundary}. After identifying an optimal observation point sampling strategy and boundary constraint method, we apply the optimal composition to estimate the unknown parameters in each test case. The results are reported in Sections \ref{1D_vibration_results} and \ref{2D_results}. 

\subsection{Observation point sampling}
\label{re:observation_points}

In our experimentation, we discovered that the collocation point distribution plays a significant role in estimating the unknown parameters in the 2D examples.

\subsubsection{2D steady state examples}
\label{re:observation_points:steady}
\begin{figure}[htbp]
\centering
\includegraphics[width=1\textwidth]{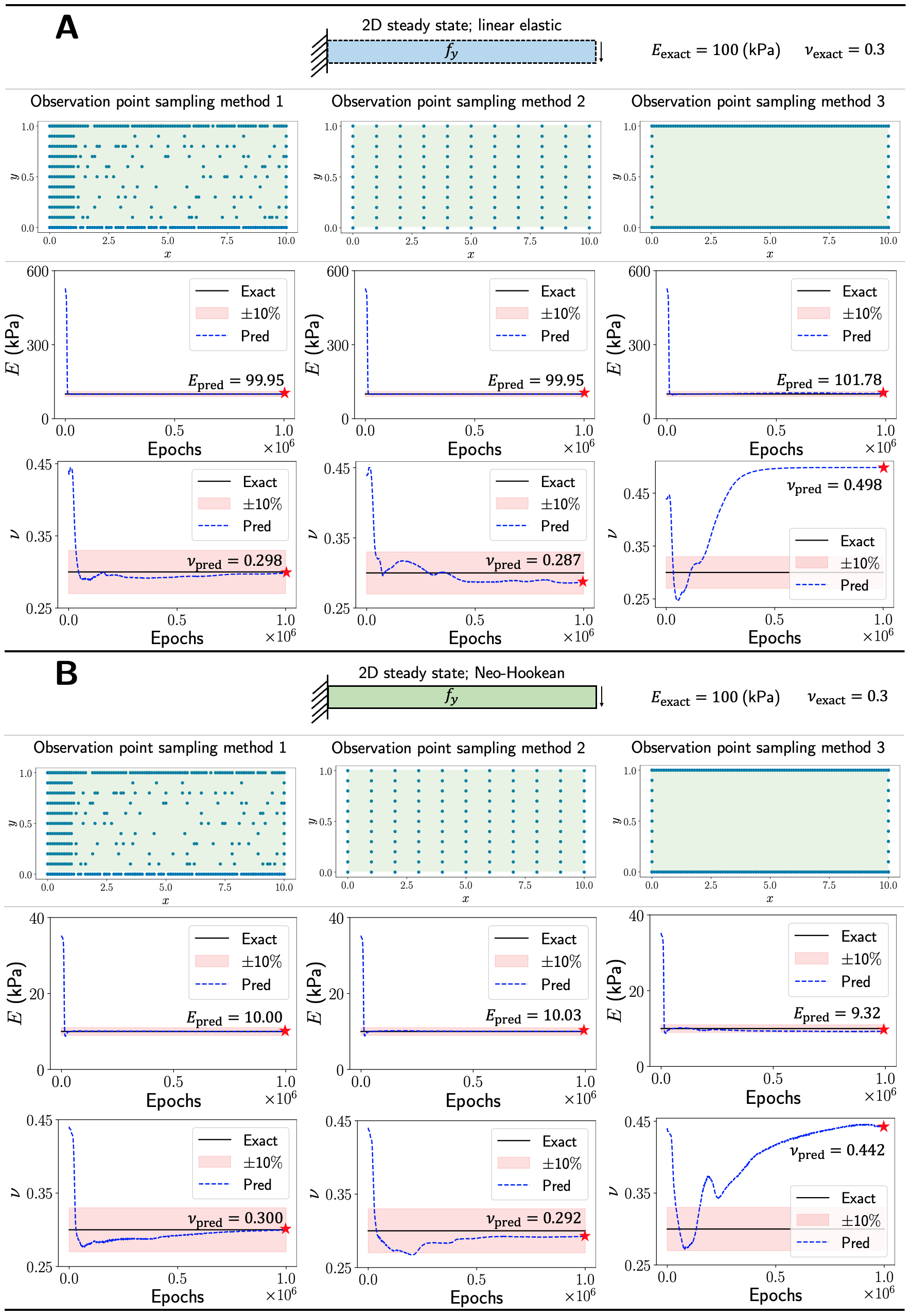}
\caption{\textbf{Observation point sampling.} Three sampling strategies are considered for the 2D linear elastic and hyperelastic steady state examples. (\textbf{A}) In the linear elastic case, the relative errors of $E_{\text{pred}}$ and $\nu_{\text{pred}}$ are $0.047\%$ and $0.539\%$ for method 1, $0.047\%$ and $4.42\%$ for method 2, and $1.800\%$ and $66.153\%$  for method 3. (\textbf{B}) In the hyperelastic case, the relative errors of $E_{\text{pred}}$ and $\nu_{\text{pred}}$ are $0.046\%$ and $0.059\%$ for method 1, $0.346\%$ and $2.554\%$ for method 2, $6.839\%$ and $47.489\%$ for method 3.}\label{point_sampling}
\end{figure}

We first perform a comparative study on the observation point sampling strategies to evaluate their influence on the unknown parameter prediction for the 2D steady state problems to identify an efficient sampling approach. We consider three methods (Fig.~\ref{point_sampling}). In method 1, we sample 111 points near the fixed boundary bounded by $x \in [0, 1]$ and $y \in [0, 1]$. In addition, we sample 191 points on the top, bottom, and right boundaries and 149 observation points in the interior (x, y) region. In method 2, we uniformly distribute 111 observation points in the spatial domain. Finally, in method 3, we sample 111 boundary points on the top and bottom edges and 11 boundary points on the left and right edges.  

The estimated unknown parameters, $E_{\text{pred}}$ and $\nu_{\text{pred}}$, are shown in Fig. \ref{point_sampling}. As shown, different observation point sampling methods do not significantly influence the convergence of $E_{\text{pred}}$. We observe that $E_{\text{pred}}$ quickly converges to the exact value in all three sampling methods for linear elastic and Neo-Hookean test examples. On the other hand, the distribution of observation points has a significant influence on $\nu_{\text{pred}}$; $\nu_{\text{pred}}$ fails to converge on method 2 (uniformly distributed observation points) and method 3 (observation points on the boundary only). This inconsistency is due to that, by Saint Venant's Principle, the Poisson's effect is insensitive in regions far from the boundaries. In the cantilever beam stress profiles, we observe concentrated stresses near the fixed end, and the stresses rapidly decay to close to zero in the free end. As such, observation points way from the fixed end may not provide enough information for the network to recover the true value of $\nu$; especially in cases where the magnitude order of $\nu$ is substantially smaller than that of $E$.

\subsubsection{2D dynamic example}
In the 2D dynamic example, we perform a comparative study to examine the accuracy of the predicted parameters with various amounts of temporal reference data. In method 1, we extract reference data from 11 time frames with $t = [0, 0.1, 0.2, ..., 0.8, 0.9, 1]$ in the FEA displacement and stress fields. In method 2, we extract reference data from 6 time frames with $t = [0, 0.1, 0.2, 0.3, 0.4, 0.5]$. Finally, in method 3, we extract reference data from 3 time frames with $t = [0, 0.1, 0.2]$. For each method, we run the inverse analysis with 5 independent networks. Each network has 3 hidden layers, with 20 neurons per layer. 

\begin{figure}[htbp]
\centering
\includegraphics[width=1\textwidth]{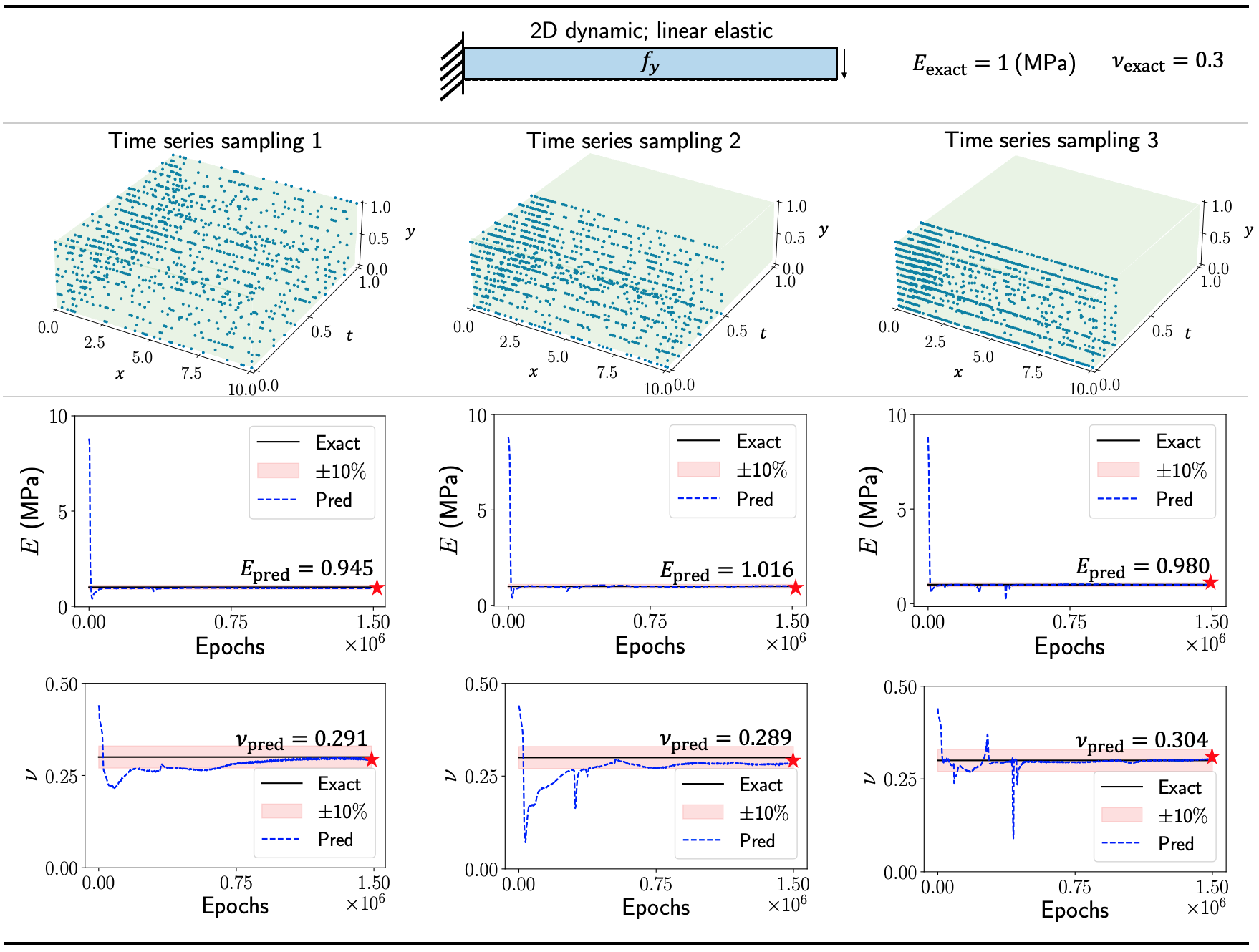}
\caption{\textbf{Time series sampling.} We consider three time series sampling strategies. In method 1, we extract 11 time frames from the FEA displacement and stress fields to use as reference data. The relative errors for $E_{\text{pred}}$ and $\nu_{\text{pred}}$ are $5.510\%$ and $2.959\%$. In method 2, we extract 6 time frames. The relative errors for $E_{\text{pred}}$ and $\nu_{\text{pred}}$ are $1.561\%$ and $5.321\%$. In method 3, we extract 3 time frames. The relative errors for $E_{\text{pred}}$ and $\nu_{\text{pred}}$ are $2.031\%$ and $1.378\%$. Our networks are able to achieve satisfactory $E_{\text{pred}}$ and $\nu_{\text{pred}}$ using reference data from as little as 3 time frames.}\label{point_sampling_2}
\end{figure}

The estimated unknown parameters, $E_{\text{pred}}$ and $\nu_{\text{pred}}$, are shown in Fig. \ref{point_sampling_2}. As demonstrated, we can achieve satisfactory $E_{\text{pred}}$ and $\nu_{\text{pred}}$ using reference data from as little as 3 time frames (the first time frame is the initial conditions). Similar to Section \ref{re:observation_points:steady}, the estimation of $E_{\text{pred}}$ is indifferent to the time series sampling method; reducing the volume of reference data in the temporal domain does not have an adverse effect on $E_{\text{pred}}$. On the contrary, our results demonstrate improved accuracy in $\nu_{\text{pred}}$ when fewer reference data are used. One reason could be that, as the volume of reference data increases, one would need sufficiently large networks to capture the interdependency between variables. However, given that the unknown parameters in our application are time-independent, including all time frames in the training process is unnecessary. As such, we find that time series sampling method 3 is an optimal choice, as it produces the most accurate predictions and is the most computationally efficient among the three methods. 

\subsubsection{Summary of observation point sampling strategy}
\begin{figure}[htbp]
\centering
\includegraphics[width=1\textwidth]{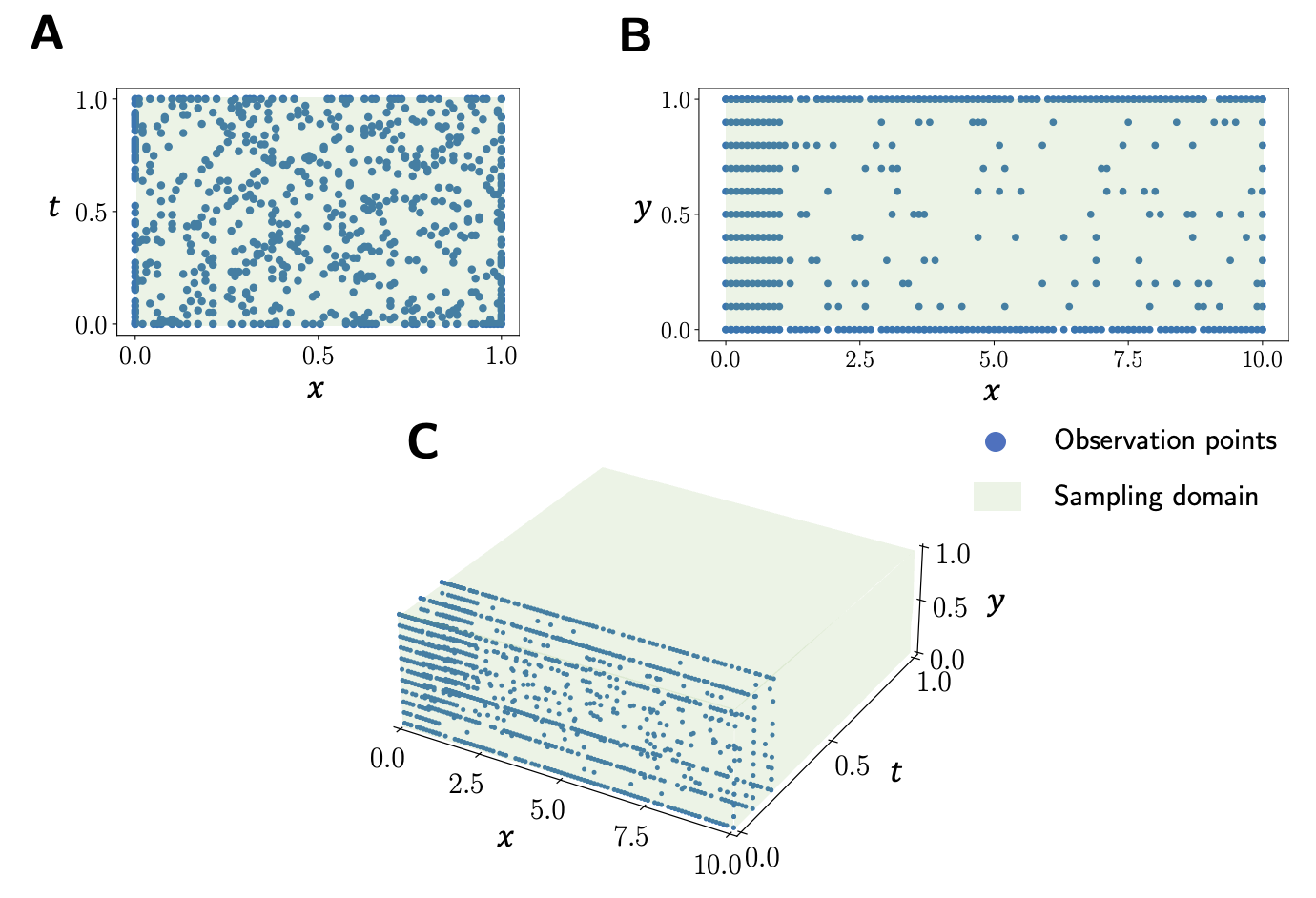}
\caption{\textbf{Observation point distribution.} (\textbf{A}) We use a total of 660 observation points for the vibration examples. Among those, 160 points are randomly distributed along the boundaries and 500 points are randomly distributed in the interior region. (\textbf{B}) We use a total of 451 observation points for the 2D steady state examples. Among those, 111 points are concentrated near the fixed boundary bounded by $x \leq 1$; 191 points are sampled on the top, bottom, and right boundaries; and 149 points are sampled in the region $x > 1$. (\textbf{C}) We follow a similar point distribution strategy as b) to extract the observation point coordinates at time instances $t = [0, 0.1, 0.2]$; a total of 1100 observation points are sampled for the dynamic example.}\label{sample_points}
\end{figure}

The observation point sampling technique used in the present work is summarized in Fig. \ref{sample_points}. For the 1D examples, we randomly distribute 160 points along the boundary bounded by $x \in [0, 1]$ and $t \in [0, 1]$ and 500 points in the interior domain. We then compute the analytical solutions given their (x, t) coordinates at the observation points. The analytical solution for longitudinal vibration is $u^* = \sin(\pi x)\cos(\pi t)$. Meanwhile, the analytical solution for lateral vibration is $u^* = \sin(\pi x)\cos(\pi^2 t)$. For the 2D steady state examples, we sample 111 points near the fixed boundary bounded by $x \in [0, 1]$ and $y \in [0, 1]$. We sample an additional 191 points randomly distributed on the top, bottom, and right boundaries. Lastly, we randomly distribute 149 observation points in the remaining (x, y) region. For the 2D dynamic example, we follow a similar sampling strategy as the steady state examples. Since both $E$ and $\nu$ are constant in time and given PINN's ability to uncover material parameters from incomplete data, we find it unnecessary to use observation points from the entirety of $t \in [0, 1]$. In the present work, the observation points are extracted at time instances $t = [0, 0.1, 0.2]$; a total of 1100 observation points are sampled. The displacements and stresses at the observation points are used as the reference data. The number of sample points is chosen arbitrarily in this work.

\subsection{Boundary constraint studies}
\label{re:boundary}
Here, we examine the unknown parameter prediction accuracy using soft and hard constraints. In soft constraints, the boundary conditions are enforced directly during the constrained optimization process by introducing a loss term in the loss function; the governing PDEs are guaranteed to satisfy the boundary conditions. Meanwhile, in the hard constraints, the boundary conditions are embedded in the PDE loss to ensure the boundary conditions are satisfied precisely (\textit{i.e.,} we apply an auxiliary function to the network outputs to enforce boundary conditions). The governing equation for this example constitutes a fourth-order spatial partial derivative and a second-order temporal partial derivative. The higher spatial and temporal derivatives in the PDEs amplify noise in the training process, which makes identifying the unknown parameter in this example a challenging task \cite{stephany2022, mattey2022}.

\subsubsection{1D lateral vibration example}
\label{re:boundary:1D}
\begin{figure}[htbp]
\centering
\includegraphics[width=1\textwidth]{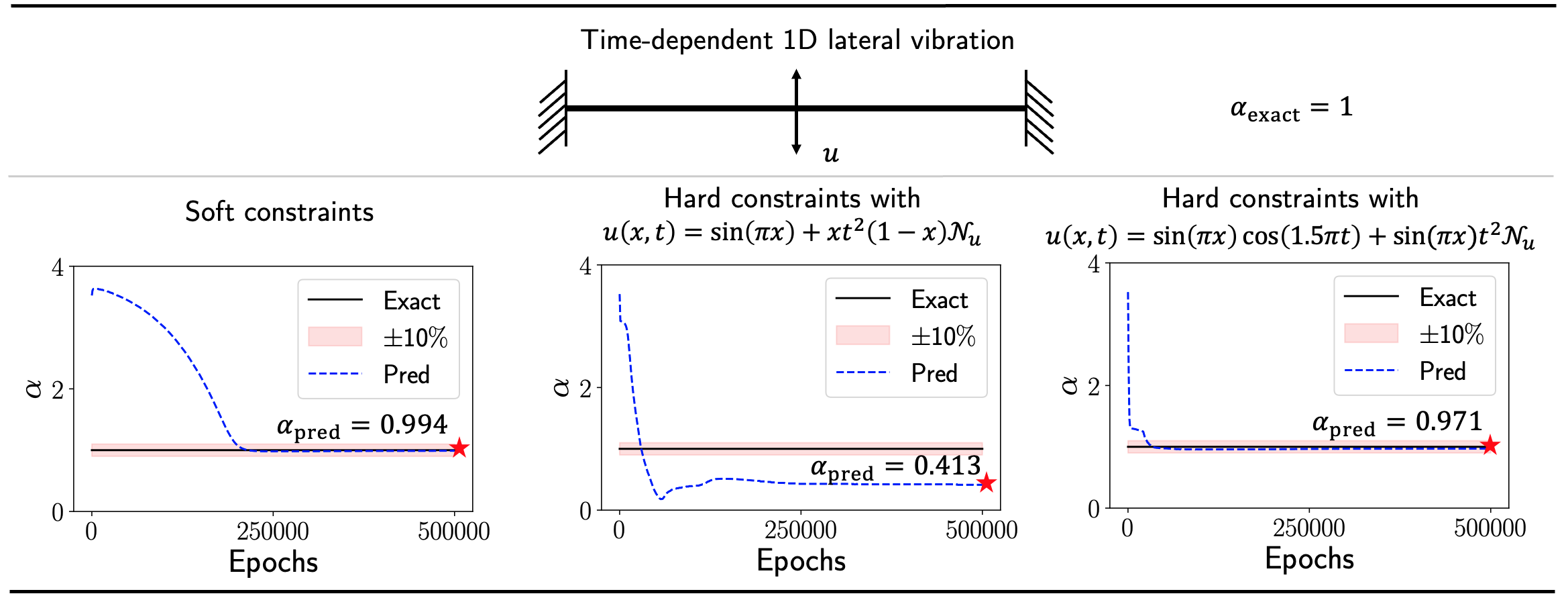}
\caption{\textbf{Boundary constraints for the 1D lateral vibration example.} We compare the accuracy of $\alpha_{\text{pred}}$ using soft constraints and hard constraints. We consider two hard constraint auxiliary functions. The results indicate that the choice of auxiliary function significantly influence the accuracy of $\alpha_{\text{pred}}$. We find that soft constraints produce the most accurate estimation of $\alpha_{\text{pred}}$, with a relative error of 0.55$\%$}\label{constraint_selection_1}
\end{figure}

In Fig. \ref{constraint_selection_1}, we compare the accuracy of the parameter $\alpha_{\text{pred}}$ using soft constraints and hard constraints. We test three different hard constraint auxiliary functions in this example; the auxiliary functions are formulated such that they satisfy the boundary conditions stated in Section \ref{govern_eqns}. Although the literature has shown that hard constraints offer better predictive power for inverse designs \cite{lu2021_2}, our results indicate that hard constraints' performance depends on the auxiliary function. Interestingly, we find that soft constraints outperform hard constraints in this particular example. However, with appropriate auxiliary functions, hard constraints can converge to an acceptable solution faster. 

\subsubsection{2D elastostatic example}
\begin{figure}[htbp]
\centering
\includegraphics[width=1\textwidth]{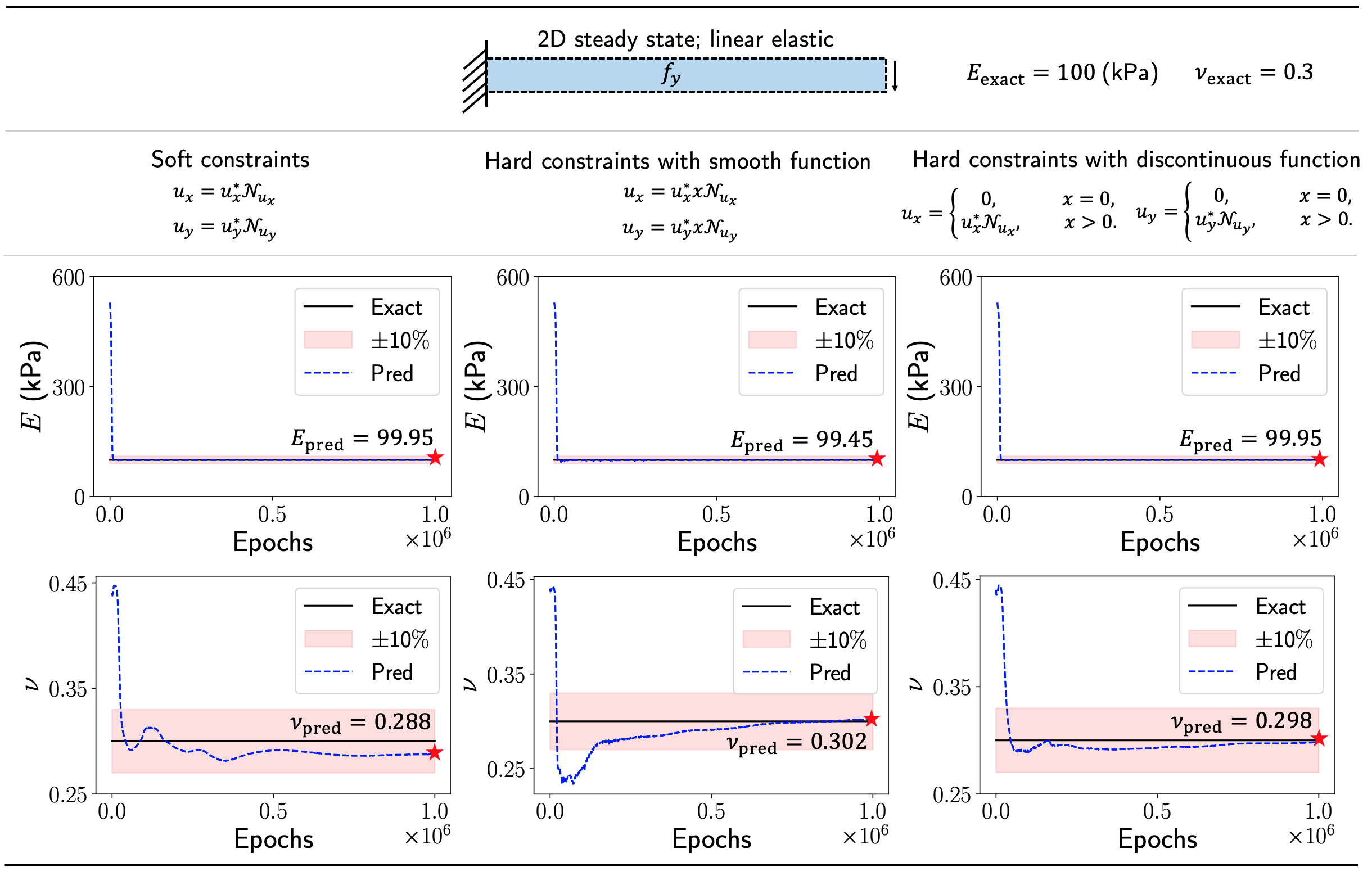}
\caption{\textbf{Boundary constraints for the 2D elastostatic example.} We examine the accuracy of $E_{\text{pred}}$ and $\nu_{\text{pred}}$ using soft constraints, hard constraints with smooth function, and hard constraints with discontinuous function. In this example, hard constraints with discontinuous function produce the best estimated $E_{\text{pred}}$ and $\nu_{\text{pred}}$, with relative errors of $0.047\%$ and $0.539\%$, respectively.}\label{constraint_selection_2}
\end{figure}

We extend the boundary constraint study to the 2D cantilever beam problem. We apply smooth and discontinuous functions to the x- and y- displacement fields for the hard constraint formulation. In this example, hard constraints with discontinuous functions offer the highest accuracy of $E_{\text{pred}}$ and $\nu_{\text{pred}}$, as shown in Fig. \ref{constraint_selection_2}. This study again highlights the importance of the auxiliary function selection on the accuracy of unknown parameters for inverse designs. 

As demonstrated in Section \ref{re:boundary:1D}, we found that the soft constraint network architecture is sufficient for accurately identifying the unknown parameter in the time-dependent 1D cases. Meanwhile, PINNs with properly chosen hard constraints \cite{lu2021_2} cast significant improvement for estimating unknown material constants in the 2D examples. 

\subsection{Parameter estimation result: time-dependent 1D examples} 
\label{1D_vibration_results}
\begin{figure}[htbp]
\centering
\includegraphics[width=1\textwidth]{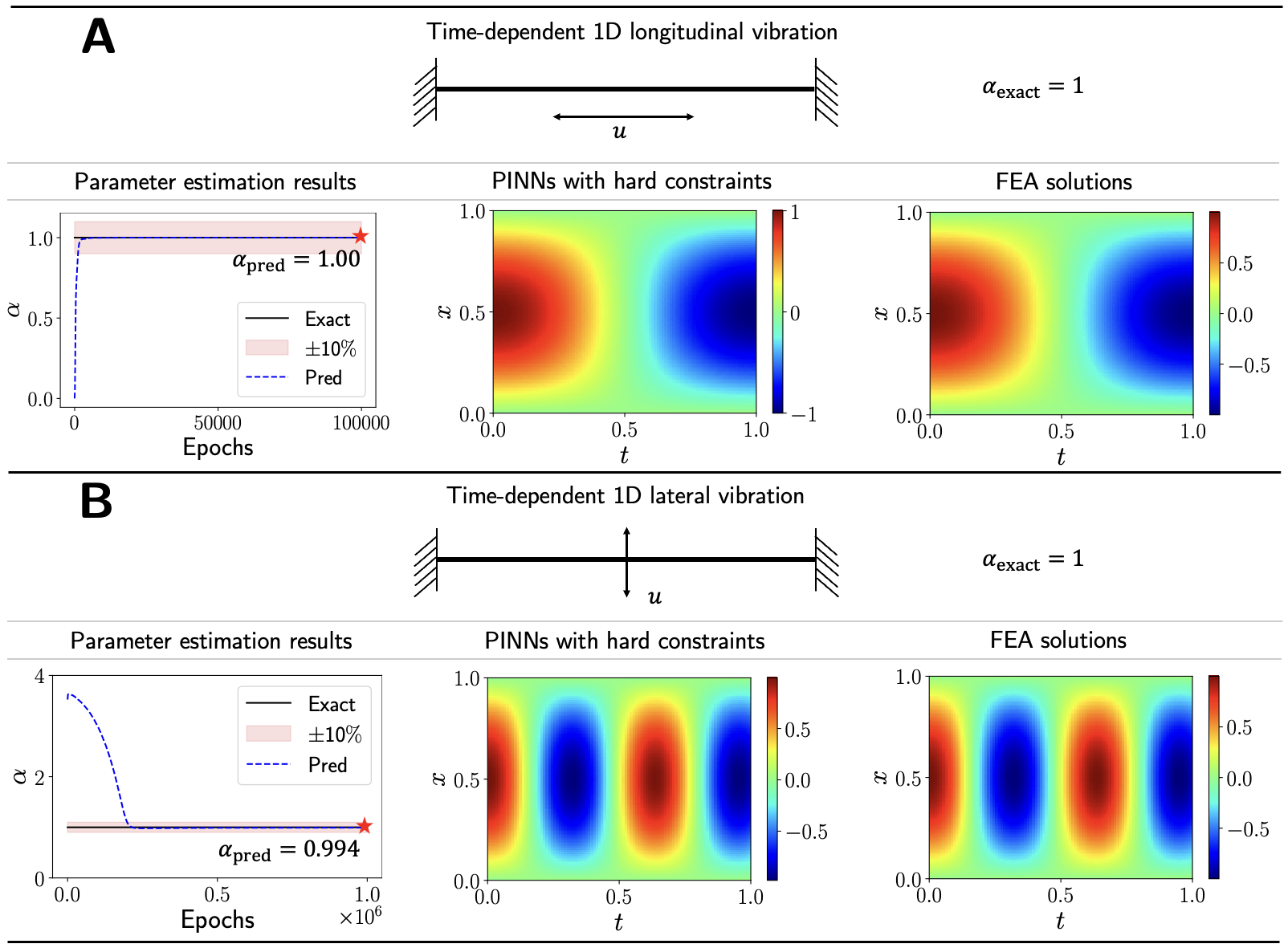}
\caption{\textbf{1D vibration parameter estimation results.} The parameter estimations, PINN predictions, and analytical solution of $u$ are provided. (\textbf{A}) The analytical solution for the longitudinal vibration example is $u_\text{analytical} = \sin{(\pi x)}\cos{(\pi t)}$. PINNs successfully recover $\alpha_{\text{pred}}$ for the time-dependent 1D longitudinal example to the true value; the relative error is 0.00\%. (\textbf{B}) The analytical solution for the lateral vibration example is $u_\text{analytical} = \sin{(\pi x)}\cos{(\pi^2 t)}$. The relative error of $\alpha_{\text{pred}}$ for the lateral vibration example is $0.55\%$. The $L_2$ relative error of the displacement fields is $2.832\%$.}\label{1dvibration}
\end{figure}

The network architecture for the 1D examples contains 3 hidden layers, with 50 neurons per layer. We set the learning rate to $10^{-3}$. In these cases, we aim to estimate the unknown parameter, $\alpha$, in the governing equations. The parameter estimation results are shown in Fig. \ref{1dvibration}. In the longitudinal vibration example, we train the network with 100000 epochs. The network estimates $\alpha_{\text{pred}}$ converges to the exact value, $\alpha_{\text{exact}} = 1$. The resulting displacement field recovers the analytical solution. In the lateral vibration example, we train the network with 500 thousand epochs. We use a hyperbolic tangent function to constrain $\alpha$ to ensure the estimated value is physical. In particular, we constrain $\alpha_{\text{pred}}$ to [0, 4] by setting $\alpha_{\text{pred}} = 2[\tanh(\hat{\alpha}_{\text{pred}})+1]$, where $\hat{\alpha}_{\text{pred}}$ is an auxiliary variable and $\alpha_{\text{pred}}$ is a scaled variable for the PDE calculation. The network estimates $\alpha_{\text{pred}} = 0.994$, with a $0.55\%$ relative error compared to the exact value, $\alpha_{\text{exact}} = 1$. The $L_2$ relative error of the resulting displacement field is $2.782\%$. 

\subsection{Parameter estimation result: 2D examples}
\label{2D_results}
\begin{figure}[htbp]
\centering
\includegraphics[width=1\textwidth]{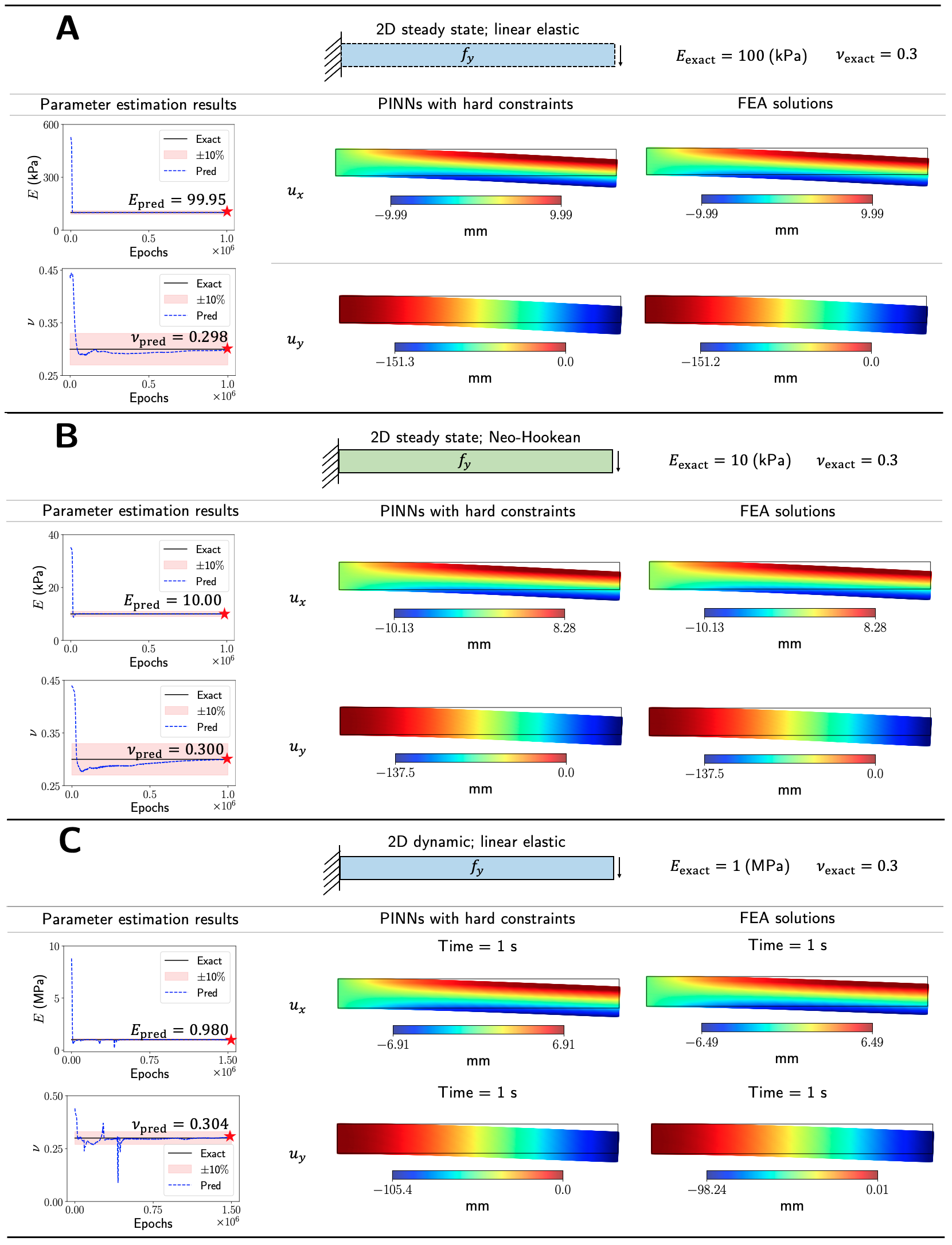}
\caption{\textbf{2D cantilever beam parameter estimation and beam displacements.} The estimated values of $E$ and $\nu$ along with the PINN and FEA approximations of $u_x$ and $u_y$ are provided. (\textbf{A}) In the 2D linear elastic steady state example, the relative errors of the estimated $E_{\text{pred}}$ and $\nu_{\text{pred}}$ are $0.047\%$ and $0.539\%$. The $L_2$ relative errors of the estimated displacement fields $u_x$ and $u_y$ are $0.049\%$ and $0.048\%$. (\textbf{B}) In the 2D hyperelastic steady state example, the relative errors of the $E_{\text{pred}}$ and $\nu_{\text{pred}}$ are $0.046\%$ and $0.059\%$. The $L_2$ relative errors of the estimated displacement fields $u_x$ and $u_y$ are $0.034\%$ and $0.034\%$. (\textbf{C}) In the 2D linear elastic dynamic example, the relative errors of the $E_{\text{pred}}$ and $\nu_{\text{pred}}$ are $2.031\%$ and $1.377\%$. The $L_2$ relative errors of the estimated displacement fields $u_x$ and $u_y$ are $3.371\%$ and $3.278\%$.} \label{2dbeam_1}
\end{figure}

\begin{figure}[htbp]
\centering
\includegraphics[width=1\textwidth]{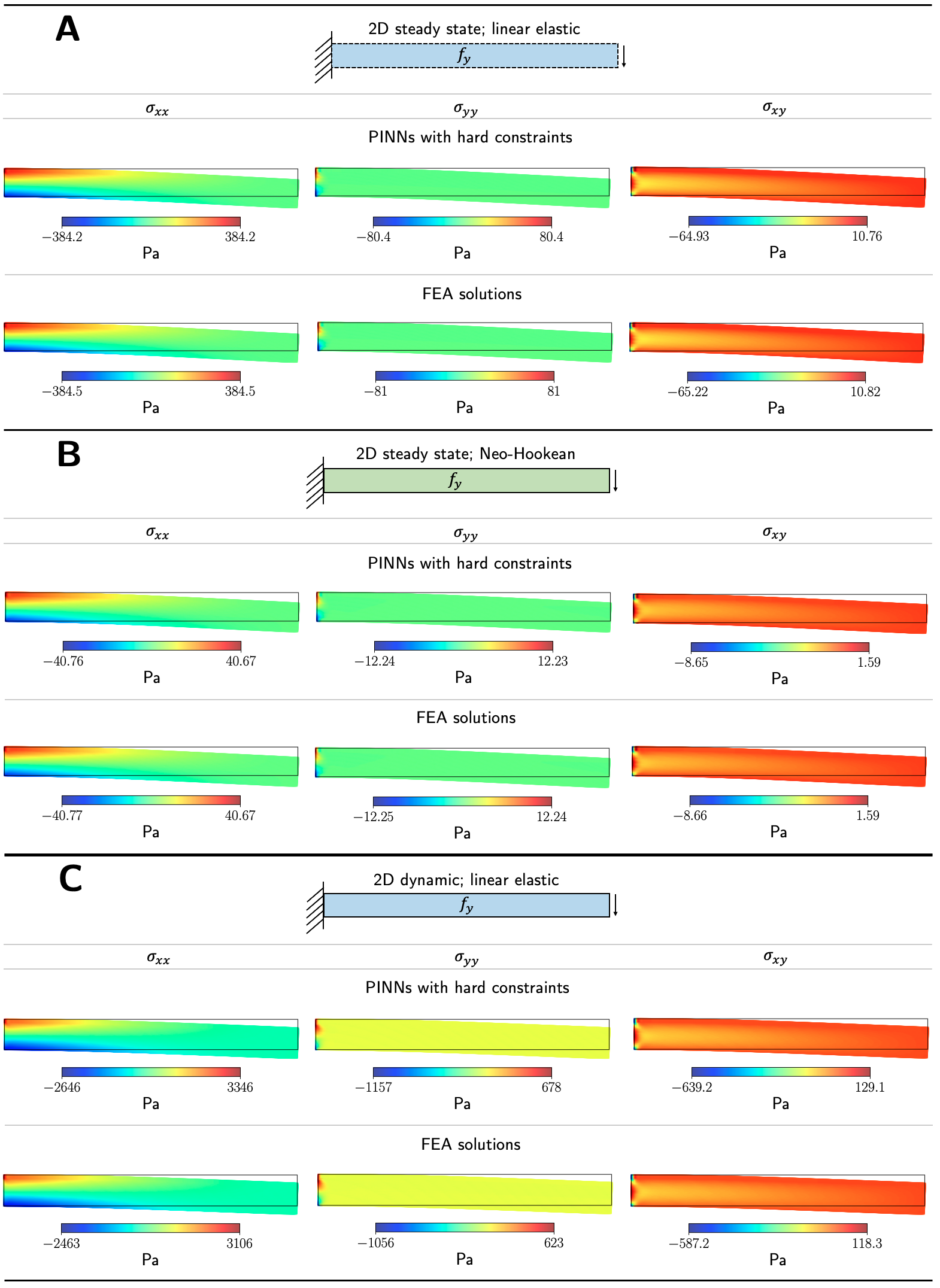}
\caption{\textbf{2D cantilever beam stresses.} We provide the PINN and FEA approximations of $\sigma_{xx}$, $\sigma_{yy}$, and $\sigma_{xy}$. (\textbf{A}) In the 2D linear elastic steady state example, the $L_2$ relative errors of the estimated stress fields $\sigma_{xx}$, $\sigma_{yy}$, and $\sigma_{xy}$ are $0.018\%$, $0.553\%$, and $0.209\%$. (\textbf{B}) In the 2D hyperelastic steady state example, the $L_2$ relative errors of the estimated stress fields $\sigma_{xx}$, $\sigma_{yy}$, and $\sigma_{xy}$ are $0.004\%$, $0.087\%$, and $0.038\%$. (\textbf{C}) In the 2D elastodynamic example, the $L_2$ relative errors of the estimated stress fields $\sigma_{xx}$, $\sigma_{yy}$, and $\sigma_{xy}$ at $t = 1$ are $3.084\%$, $3.388\%$, and $3.832\%$.}\label{2dbeam_2}
\end{figure}

We use five independent neural networks in the 2D examples. For the steady state examples, each network has 3 hidden layers, with 15 neurons per layer. For the dynamic example, we use 3 hidden layers per network, with 20 neurons per layer. The performance of various network architectures are presented in Section~\ref{network variation}. We set the learning rate to $10^{-3}$. In these cases, we aim to estimate the unknown Young's modulus and Poisson's ratio, $E$ and $\nu$, in the material constitutive laws. The steady state models are trained with 1 million epochs. Meanwhile, the dynamic model is trained with 1.5 million epochs. We use double precision floating point in the 2D examples. For all three examples, we compute the reference displacements and Cauchy stresses using an open-source finite element software, FEniCS \cite{Anders2010}, based on pre-defined $E_{\text{exact}}$ and $\nu_{\text{exact}}$ values. The approximate $E_{\text{pred}}$ and $\nu_{\text{pred}}$, as well as the displacement and stress fields, are shown in Figs. \ref{2dbeam_1} and \ref{2dbeam_2}. The relative errors of $E_{\text{pred}}$, $\nu_{\text{pred}}$, and the resulting $u_x$, $u_y$, $\sigma_{xx}$, $\sigma_{yy}$, and $\sigma_{xy}$ are summarized in Table  \ref{L2errors}.

The differences in the magnitude order between the displacement and stress fields can pose a challenge in the training process and influence solution convergence. In the present work, the displacements are ranging from $\mathcal{O}(10^{-1})$~m to $\mathcal{O}(10^{-3})$~m, while the stresses are ranging from $\mathcal{O}(1)$~Pa to $\mathcal{O}(10^{2})$~Pa. To improve convergence, we rescale the network output displacements and stresses by their respective maximum absolute values in the reference solution. This helps ensure that the network output variables are all in $\mathcal{O}(1)$. Further, the magnitude order disparity between material parameters, $E$ and $\nu$, can also present difficulties in identifying an accurate solution. Similar to the 1D examples (Section \ref{1D_vibration_results}), we use a $\tanh$ function to not only ensure the predicted material constants are in a realistic range but also keep the auxiliary variables predicted by the network in similar magnitude orders. 

\begin{figure}[htbp]
\centering
\includegraphics[width=1\textwidth]{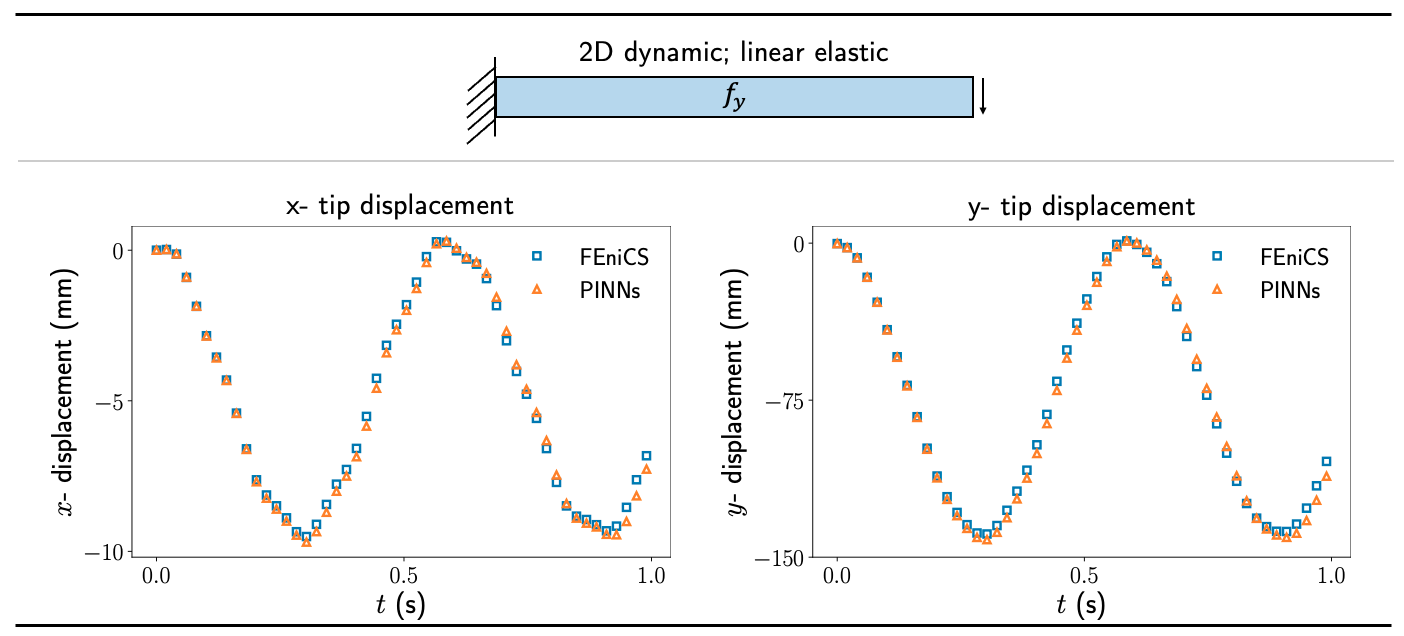}
\caption{\textbf{2D cantilever beam tip deflection evolution.} We compare the displacement at the cantilever tip against FEniCs for the elastodynamic example. The displacements predicted by PINNs agree with the FEA solution qualitatively, with the PINN predicted displacements exhibiting slightly higher dissipation as time evolves. The $L_2$ relative errors of the estimated x- and y- tip displacements for time $t\in$ [0, 1] are $0.346\%$ and $3.306\%$, respectively. }\label{2dbeam_3}
\end{figure}

In the 2D linear elastic steady state example, we constrain $E_{\text{pred}}$ to [0, 700]~kPa and $\nu$ to [0, 0.5]. The network estimates $E_{\text{pred}} = 99.95$~kPa and $\nu_{\text{pred}}=0.298$, compared to the exact values $E_{\text{exact}} = 100$~kPa and $\nu_{\text{exact}} = 0.3$. In the 2D hyperelastic steady state example. we constrain $E_{pred}$ to [0, 60]~kPa and $\nu$ to [0, 0.5]. The exact values of Young's modulus and Poisson's ratios are $E_{\text{exact}} = 10$~kPa and $\nu_{\text{exact}} = 0.3$. The predicted values are $E_{\text{pred}} = 10.0$~kPa and $\nu_{\text{pred}}=0.300$. In the 2D dynamic example, we constrain $E_{\text{pred}}$ to [0, 10]~MPa and $\nu$ to [0, 0.5]. The exact values of Young's modulus and Poisson's ratios are $E_{\text{exact}} = 1$~MPa and $\nu_{\text{exact}} = 0.3$. The predicted values are $E_{\text{pred}} = 0.980$~MPa and $\nu_{\text{pred}}=0.304$. The x- and y- displacement at the free end of the cantilever for $t\in[0, 1]$ are presented in Fig. \ref{2dbeam_3}. The PINN approximate displacements show excellent alignment with reference data generated from FEniCS. The $L_2$ relative errors for the x- and y- tip displacements are $3.462\%$ and $3.306\%$. \\

\begin{table}[htbp]
 \caption{\textbf{Relative errors of the unknown parameters, $E_{\text{pred}}$ and $\nu_{\text{pred}}$, and the resulting displacement and stress fields.} (\textbf{A} and \textbf{B}) The relative errors of both the estimated material constants and mechanical quantities are well under $1\%$ for the steady state examples. (\textbf{C}) For the dynamic example, the relative errors of the material parameters and mechanical quantities are under $2.5\%$ and $4\%$, respectively.} \label{L2errors}
 \begin{tabular}{c}
 \includegraphics[width=1\textwidth]{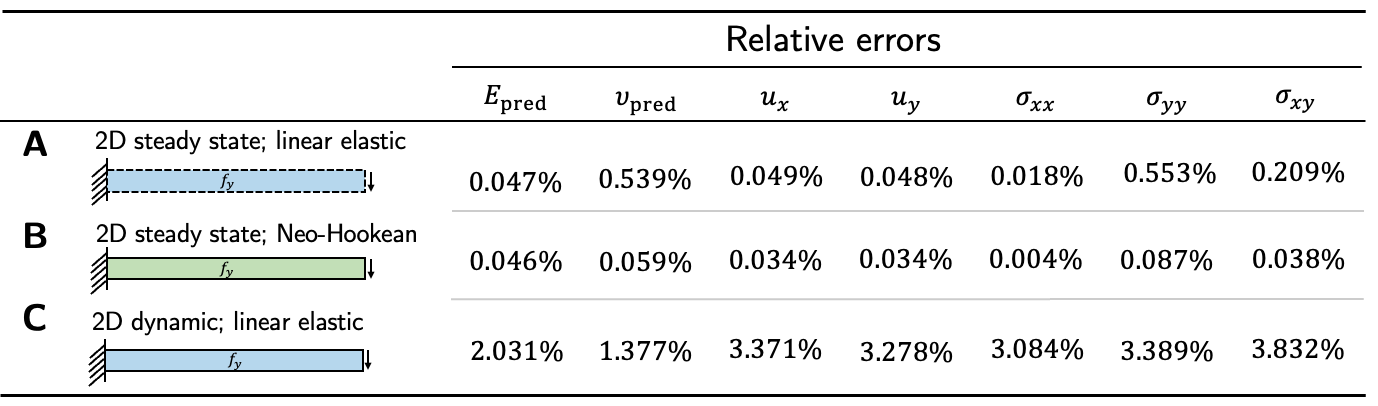}
 \end{tabular}
\end{table}

\section{Discussion}
\label{discussion}
We present a generalized approach for PINNs to solve inverse problems in solid mechanics. Traditionally, inverse FEA is a popular choice for solving inverse problems. However, the convergence of inverse FEA is highly dependent on the mesh quality and measured data. Although the application of PINNs for solving the inverse problem in solid mechanics is still in its infancy, the neural network approach for inverse problems has been demonstrated to have advantages, such as its insensitivity to noisy and incomplete data \cite{berg2021}. Furthermore, as shown in the present work, this novel approach can produce highly accurate estimations of material parameters in the linear elastic and hyperelastic domains in both steady state and dynamic situations. As such, PINN has substantial potential for application in diverse fields dependent upon solid mechanics and biomechanics.

Although publications on PINNs have grown exponentially since Raissi's publication in 2019 \cite{raissi2019}, previous work on applying PINNs for solving inverse problems in continuum solid mechanics is sparse, especially for 2D or 3D problems. Researchers recently have applied the weak form of conservation equations to identify material properties \cite{Thakolkaran2022}. However, the weak form requires a denser mesh than the strong form to obtain accurate integral estimations and parameter predictions. This limitation will result in higher computational costs as the complexity of the problem increases. At the time of writing, there are only a few research articles that describe a framework for material identification in the linear elastic and hyperelastic domains, utilizing the strong form of the conservation laws \cite{haghighat2021, zhang2022}. This is partially due to the challenges of obtaining satisfactory parameter estimations for inverse problems in most practical engineering applications. In many realistic solid mechanics problems, the mechanical quantities (\textit{i.e.,} the displacement and stress fields) as well as the material constants (\textit{e.g.,} Young's modulus, $E$, and the Poisson's ratio, $\nu$) are often differed by multiple orders of magnitude. The displacement fields are commonly tiny, in the present work ranging from $\mathcal{O}(10^{-1})$~m to $\mathcal{O}(10^{-3})$~m, compared to the magnitude of the stress fields, in the present work ranging from $\mathcal{O}(1)$~Pa to $\mathcal{O}(10^{2})$ Pa. With such a small displacement magnitude, the loss term $\mathcal{L}_{\text{Data}}$ for the displacement fields becomes insensitive to the deviations of network parameters, $\theta_{\text{NN}}$. In addition, the magnitude of the material parameter, $E$, is commonly many orders higher than the material parameter, $\nu$. This huge disparity between $E$ and $\nu$ present difficulties in identifying an accurate solution for $\nu$ as $E$ dominates the mechanical response mathematically. Because of the preceding reasons, a pre-trained network is used to estimate the unknown parameters of interest in the work of \cite{haghighat2021, zhang2022}.

Using a pre-trained network may reduce training time for problems of similar variants. Otherwise, it may be less helpful. The inversion examples in the present work are performed de novo without reliance on a pre-trained network; this highlights the inherent generalizability of our framework. We mitigate the challenges described above through the following steps. First, we use an independent network for each output variable, as suggested in \cite{haghighat2021}. Second, we determine appropriate observation point sampling strategies for the problems of interest to ensure the influence of all material parameters is captured (as demonstrated in the cantilever beam example, the Poisson's effect is insensitive in regions away from the fixed boundary, which contributes to the difficulties of approximating the unknown parameter, $\nu$). Third, we reduce the weight of PDE loss to increase the influence of data in the training process. Fourth, we transform each network output variable by multiplying its corresponding maximum absolute value in reference data and apply hard constraints if needed. In our experience, the previous four steps are sufficient for identifying the material parameters in most cases. In situations where the network fails to converge due to considerable differences between $E$ and $\nu$, for example, in our 2D linear elastic dynamic case, we reduce the magnitude order of $E$ in the approximation process to improve the influence of $\nu$ in the estimated stress fields. In linear elastic problems, the material parameter, $E$, and the displacement fields are inversely proportional. That means, reducing the magnitude order of $E$ will increase the same magnitude order in the predicted displacement fields. As such, appropriate transformation of the predicted displacement fields is applied.

Further, PINNs hold several benefits over traditional engineering methods. Traditional numerical methods, such as finite element and finite volume, typically rely on complex spatial and temporal discretization schemes that could easily result in thousands of lines of code. In addition, the solution accuracy in classic numerical methods strongly depends on the mesh quality and element formulation. Numerical solutions from mesh-based finite element and finite volume methods are highly susceptible to numerical instability when handling complex geometry due to element distortion. Unlike classic mesh-based methods, PINNs are mesh-free, which eliminates element-related challenges. In addition, PINNs work directly with the strong form of conservation equations. The partial derivatives in the governing equation are computed using automatic differentiation, bypassing the need for numerical discretization schemes. Furthermore, the development of high-level deep learning libraries such as DeepXDE~\cite{lu2021} has allowed PINN frameworks to be easily set up in less than one hundred lines of code. These simple-to-use and user-friendly libraries drastically reduces the time needed to build and apply algorithms for inverse analyses. Finally, unlike traditional numerical methods, in which the parameter search process starts from scratch in every new inverse analysis, the PINN network parameters can be stored and reused when solving similar problems to improve network training time and solution accuracy. 

Potential applications for determining material properties using PINNs are diverse. Computational modeling of the physical behavior of biological tissues has significant potential to inform patient-specific medicine \cite{freutel2014, lin2022, liu2019}. For example, the \textit{in silico} modeling of cardiac valves has the potential to allow the optimization of valve repair techniques before actual application of the repair in a patient \cite{narang2021, Lee2019, Wu2022}. However, accurate results will depend on knowledge of the material properties of the valve leaflets, which may vary across age, valve type, and specific pathology. The capability of PINNs to determine physical parameters even in the setting of missing or noisy data makes it well suited to extracting material properties from clinical medical imaging (3D ultrasound, computed tomography, magnetic resonance imaging) of individual patients, which in turn will facilitate the application of precision medicine based on computational models. This example generalizes to many applications common in biological systems. The combination of sparse data and a physical framework can be leveraged to answer questions where traditional approaches may not be feasible or even capable of generating a solution.

\section{Conclusion}
\label{conclusion}

We have described the development and use of physics-informed neural networks (PINNs) to solve inverse problems in solid mechanics. Additionally, five classic inverse problems have been solved using PINNs with a demonstrated accuracy of the material parameters in the order of $1\%$. This provides proof of concept that our PINN framework will work for material parameter estimation. Future work will utilize PINNs in multiple domains, including complex biological systems in medicine. 

\section{Acknowledgment}
This work was funded by the Cora Topolewski Cardiac Research Fund At the Children's Hospital of Philadelphia (CHOP), The Pediatric Valve Center Frontier Program at CHOP, a Additional Ventures Single Ventricle Research Fund Expansion Award, and the National Institutes of Health (USA). WW was supported by NHLBI T32 HL007915 and NIH R01 HL153166. MAJ was supported by NIH R01 HL153166. MD and LL were supported by the U.S. Department of Energy [DE-SC0022953].

\appendix

\section{Additional details on the test examples} \label{govern_eqns}

\subsection{1D longitudinal vibration}

\begin{figure}[h!]
\centering
\includegraphics[width=0.4\textwidth]{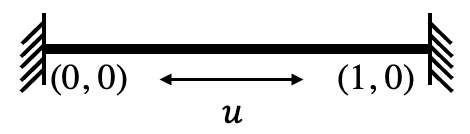}
\caption{\textbf{Longitudinal vibration.} The beam is fixed on both ends. An initial longitudinal vibration, $u(x,0)=\sin(\pi x)$, is applied to the beam.}\label{1D_long}
\end{figure}

In the longitudinal vibration example, a beam with length $L=1$ is subjected to an initial longitudinal vibration $u$ as shown in Fig. \ref{1D_long}. 

The mathematical model of this example can be characterized by the wave equation: 
\begin{equation*} 
\frac{\partial^2 u}{\partial t^2}= \alpha^2\frac{\partial^2 u}{\partial x^2}, 
\end{equation*}
with the following boundary conditions and initial conditions:
\begin{gather*}
u(0, t) = u(1, t) = 0, \quad \textrm {for} \quad 0\leq t \leq 1; \\
u(x, 0) = \sin(\pi x), \quad \frac{\partial u(x, 0)}{\partial t} = 0, \quad \textrm {for} \quad 0\leq x \leq 1.
\end{gather*}
The unknown variable, $\alpha$, is set to 1. The analytical solution of this example is: 
\begin{equation*} 
u^*=\sin(\pi x)\cos(\pi t). 
\end{equation*}

In this example, the weights in the loss function, $w_i$ for $i \in$ [IC, BC, PDEs, and reference data], are set to 1.

\subsection{1D lateral vibration}

\begin{figure}[h!]
\centering
\includegraphics[width=0.4\textwidth]{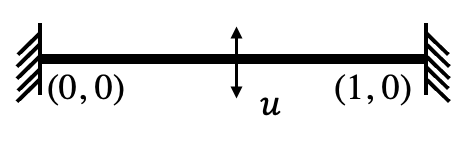}
\caption{\textbf{Lateral vibration.} The beam of length $L=1$ is fixed on both ends. We apply an initial lateral vibration, $u(x, 0)=\sin(\pi x)$, to inject lateral vibration to the system.}\label{1D_lat}
\end{figure}

In the second example, an Euler-Bernoulli beam with length $L=1$ is subjected to an initial lateral displacement $u$ as shown in Fig. \ref{1D_lat}. 

The Euler-Bernoulli beam equation of the lateral vibration system shown above is expressed as: 
\begin{equation}\label{euler-bernoulli}
\frac{\partial^2u}{\partial t^2}= -\alpha^2\frac{\partial^4u}{\partial x^4}, 
\end{equation}
with the following boundary conditions and initial conditions:
\begin{gather*}
u(0, t) = u(1, t) = 0, \quad \textrm {for} \quad 0\leq t \leq 1; \\
u(x, 0) = \sin(\pi x), \quad \frac{\partial u(x, 0)}{\partial t} = 0, \quad \textrm {for} \quad 0\leq x \leq 1;\\
\frac{\partial^2u(0, t)}{\partial x^2} = \frac{\partial^2u(0, t)}{\partial x^2} = 0, \quad \textrm {for} \quad 0\leq t \leq 1.
\end{gather*}


Similar to the 1D lateral vibration example, we set $\alpha$ to 1. The analytical solution of Eq. \eqref{euler-bernoulli} becomes:
\begin{equation*} 
u^*=\sin(\pi x)\cos(\pi^2 t). 
\end{equation*}

In this example, the weights in the loss function, $w_i$ for $i \in$ [IC, BC, PDEs, and reference data], are also set to 1.

\subsection{2D linear elastic cantilever beam}

\begin{figure}[h!]
\centering
\includegraphics[width=0.4\textwidth]{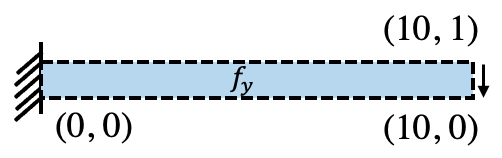}
\caption{\textbf{2D linear elastic cantilever beam.} The linear elastic cantilever beam with a dimension of $10$~m $\times$ $1$~m is subjected to a downward body force $f_y=1$~N.}\label{2D_linear_elastic}
\end{figure}

In the third example, we consider a $10$~m $\times$ $1$~m cantilever beam fixed on the left end as shown in Fig. \ref{2D_linear_elastic}. The beam is made of linear elastic material and subjected to a downward body force $f_y = 1$~N. 

In this steady state example, we assume plane-stress formulation. The momentum balance equation is expressed as:
\begin{equation*}
\sigma_{ij,j}+f_i= 0.
\end{equation*}
The isotropic linear elastic material constitutive model is defined as:
\begin{equation*}
\mathbf{\sigma} = \mathbf{C} \cdot \mathbf{\epsilon},
\end{equation*}
with 
\begin{gather*}
\mathbf{\sigma} = 
\begin{bmatrix}
\sigma_{xx} \\ \sigma_{yy} \\ \sigma_{xy}
\end{bmatrix}; \quad 
\mathbf{C} = \frac{E}{(1-\nu^2)}
\begin{bmatrix}
1 & \nu & 0 \\ 
\nu & 1 & 0 \\ 
0 & 0 & (1-\nu)
\end{bmatrix};
\quad
\mathbf{\epsilon} = 
\begin{bmatrix}
\epsilon_{xx} \\ \epsilon_{yy} \\ \epsilon_{xy}
\end{bmatrix}
\end{gather*}
The kinematic relation is:
\begin{gather*}
\epsilon_{xx} = \frac{\partial u_x}{\partial x}; \quad \epsilon_{yy} = \frac{\partial u_y}{\partial y}; \quad
\epsilon_{xy} = \frac{1}{2}[\frac{\partial u_x}{\partial y}+\frac{\partial u_y}{\partial x}].
\end{gather*} 
 
Young's modulus and Poisson's ratio of this example are $10^5$~Pa and 0.3, respectively. The weights in the loss function, $w_{\text{PDEs}}$ and $w_{\text{Data}}$, are set to $1e^{-10}$ and 1, respectively.

\subsection{2D hyperelastic cantilever beam}

\begin{figure}[h!]
\centering
\includegraphics[width=0.4\textwidth]{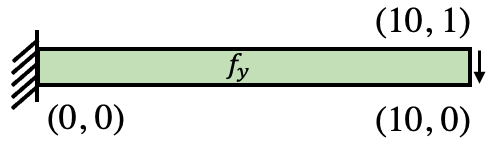}
\caption{\textbf{2D hyperelastic cantilever beam.} The $10$~m $\times$ $1$~m hyperelastic cantilever beam is subjected to a body force $f_y=0.1$~N.}\label{2D_neohookean}
\end{figure}

In the fourth example shown in Fig. \ref{2D_neohookean}, we consider the same geometry as the third example but with Neo-Hookean material. 

The constitutive model for compressible isotropic hyperelastic material is expressed as:
\begin{equation*} 
\sigma_{ij} = \frac{1}{J}P_{ik}F^T_{kj},
\end{equation*}
where $F$ is the deformation gradient tensor defined as $F_{ij} = \delta_{ij}+u_{i,j}$, $J=det(F)$, and P is the first Piola-Kirchhoff stress tensor. The first Piola-Kirchhoff stress for compressible Neo-Hookean material is as follows:
\begin{equation*} 
P = \frac{\partial \Psi}{\partial F} = \mu F+[\lambda log(J)-\mu]F^{-T},
\end{equation*}
where $\lambda$ and $\mu$ are the Lam\'e's elasticity parameters. Further, $\Psi$ is the strain energy density function. Assuming plane-strain formulation:
\begin{gather*}
\lambda = \frac{E\nu}{(1+\nu)(1-2\nu)}, \quad \textrm{and} \quad \mu = \frac{E}{2(1+\nu)}. 
\end{gather*}
The Neo-Hookean strain energy density function, $\Psi$, is expressed as:
\begin{equation*} 
\Psi(I_1, J) = \frac{1}{2}\lambda[log(J)]^2-\mu log(J)+ \frac{1}{2}\mu (I_1-2),
\end{equation*}
where $I_1 = trace(F^T\cdot F)$. The weights in the loss function, $w_{\text{PDEs}}$ and $w_{\text{data}}$, are set to $10^{-8}$ and 1, respectively.

\subsection{2D dynamic cantilever beam}

In the fifth example, we extend the 2D linear elastic cantilever beam example to dynamic analysis. The beam has density $\rho = 1 \frac{kg}{m^3}$. Young's modulus and Poisson's ratio are $10^{6}$~Pa and 0.3, respectively. The applied body force in this example is $f_y= 5$~N. The momentum balance equation becomes:
\begin{equation*}
\sigma_{ij,j}+f_i= \rho \partial_{tt}u_i, 
\end{equation*}
Consider plane-strain formulation, the $\mathbf{C}$ matrix in the constitutive relation $\mathbf{\sigma} = \mathbf{C} \cdot \epsilon$ is 
\begin{gather*}
\mathbf{C} = 
\frac{E}{(1+\nu)(1-2\nu)}
\begin{bmatrix}
(1-\nu) & \nu & 0 \\ 
\nu & (1-\nu) & 0 \\ 
0 & 0 & (1-2\nu)
\end{bmatrix}.
\end{gather*}

The weights in the loss function, $w_{\text{PDEs}}$ and $w_{\text{data}}$, are set to $10^{-8}$ and 1, respectively.

\section{FEA analysis} \label{verfication}

We used FEniCS to generate the displacement and stress reference data for the 2D examples. The cantilever beam geometry is discretized into 1000 second-order rectangular elements with $\Delta x = \Delta y = 0.1$~m.  For the linear elastic static example, we validate the maximal deflection obtained from FEA against the analytical solution from Euler-Bernoulli beam theory: $u_{y, max}= \frac{\rho g L^4}{EI}$. The maximum deflection from the analytical solution and FEA are $-0.15$ ~m and $-0.151$ ~m, respectively. The relative error is approximately $0.67\%$. In the 2D dynamic examples, a Newmark implicit time integration scheme is used to facilitate the evolution of displacement and stress in time. We validate the frequency of the y- tip displacement against the first natural frequency for the cantilever beam derived from Euler-Bernoulli beam theory: $f = \frac{1.875^2}{2\pi L^2}\sqrt{\frac{EI}{\rho A}}$. The natural frequency from the analytical solution is 1.61~Hz, meanwhile, the natural frequency from the FEA solution is 1.66~Hz. The relative error is approximately $3\%$.

\section{Variation of network architectures}\label{network variation}

\begin{figure}[h!]
\centering
\includegraphics[width=1\textwidth]{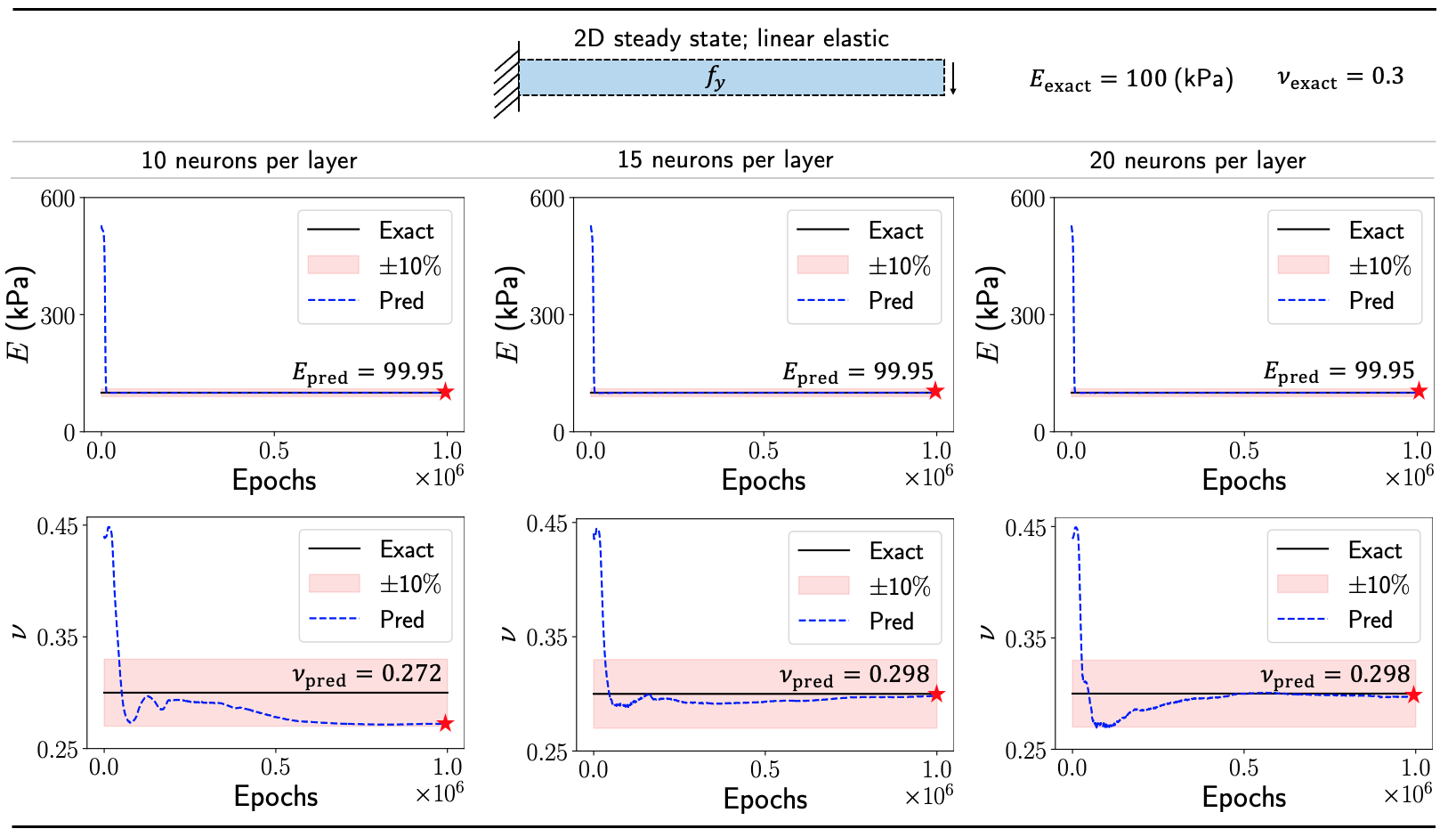}
\caption{\textbf{Parametric study of the network size on prediction accuracy.} We compare the relative errors of $E_{\text{pred}}$ and $\nu_{\text{pred}}$ for three different network sizes. In the first network architecture (5-3-10), the relative errors of $E_{\text{pred}}$ and $\nu_{\text{pred}}$ are $0.047\%$ and $7.04\%$. In the second network architecture (5-3-15), the relative errors of $E_{\text{pred}}$ and $\nu_{\text{pred}}$ are $0.047\%$ and $0.18\%$. In the third network architecture (5-3-20), the relative errors of $E_{\text{pred}}$ and $\nu_{\text{pred}}$ are $0.21\%$ and $2.56\%$.}\label{network_study}
\end{figure}

In this section, we study the influence of network architectures on PINNs' predictive capability. In our work, we use 5 independent networks to estimate $E_{\text{pred}}$ and $\nu_{\text{pred}}$ in the 2D examples. Each independent network has 3 layers of neurons; the same number of neurons are applied to each layer. Here, we variate the width of the network by adjusting the number of neurons to 10, 15, and 20 neurons per layer. We abbreviate the three network architectures to 5-3-10, 5-3-15, and 5-3-20, wherein the first number refers to the number of independent networks, the second number refers to the number of layers per network, and the last number refers to the number of neurons per layer. The resulting $E_{\text{pred}}$ and $\nu_{\text{pred}}$ are presented in Fig. \ref{network_study}. As shown, the width of the neural network does not have a significant influence on the convergence of $E_{\text{pred}}$; $E_{\text{pred}}$ converges rapidly to the ground truth in less than 100K epochs. As we increase the number of neurons from 10 to 15, the relative error reduces from $7.04\%$ to $0.18\%$. However, $\nu_{\text{pred}}$ starts to deviate from the ground truth as we keep increasing the width of the network to 20 neurons per layer. This indicates that a network width of 15 is most suitable for the examples at hand.  

\bibliographystyle{unsrt}  
\bibliography{library}

\end{document}